\documentclass[preprint2]{aastex6}

\newcommand{\spitzer}{{\it Spitzer}}
\newcommand{\herschel}{{\it Herschel}}
\newcommand{\hst}{{\it HST}}

\newcommand{\hi}{\ion{H}{1}}
\newcommand{\hii}{\ion{H}{2}}

\newcommand{\ariii}{[\ion{Ar}{3}]}
\newcommand{\nii}{[\ion{N}{2}]}
\newcommand{\neii}{[\ion{Ne}{2}]}
\newcommand{\neiii}{[\ion{Ne}{3}]}
\newcommand{\oii}{[\ion{O}{2}]}
\newcommand{\oiii}{[\ion{O}{3}]}
\newcommand{\sii}{[\ion{S}{2}]}
\newcommand{\siii}{[\ion{S}{3}]}
\newcommand{\siv}{[\ion{S}{4}]}

\begin{document}

\title{Infrared Galaxies in the Field of the Massive Cluster Abell~S1063:
Discovery of a Luminous Kiloparsec-Sized \ion{H}{2} Region\\ in a
Gravitationally Lensed IR-Luminous Galaxy at $z=0.6$
}

\author{
Gregory L. Walth\altaffilmark{1,2,3},
Eiichi Egami\altaffilmark{3},
Benjamin Cl\'ement\altaffilmark{4},
Timothy D. Rawle\altaffilmark{5},
Marie Rex\altaffilmark{3},
Johan Richard\altaffilmark{4},
Pablo P{\'e}rez-Gonz{\'a}lez\altaffilmark{6,7},
Fr{\'e}d{\'e}ric Boone\altaffilmark{8},
Miroslava Dessauges-Zavadsky\altaffilmark{9},
Jeff Portouw\altaffilmark{3},
Benjamin Weiner\altaffilmark{3},
Ian McGreer\altaffilmark{3},
Evan Schneider\altaffilmark{10}
}


\altaffiltext{1}{The Observatories of the Carnegie Institution for Science,
813 Santa Barbara St., Pasadena, CA 91101, USA; gwalth@carnegiescience.edu}
\altaffiltext{2}{UC San Diego, Center for Astrophysics \& Space Sciences,
9500 Gilman Drive, La Jolla, CA, USA}
\altaffiltext{3}{Steward Observatory, University of Arizona, 933 N. Cherry Ave,
Tucson, AZ 85721, USA}
\altaffiltext{4}{Univ Lyon, Univ Lyon1, Ens de Lyon, CNRS, Centre de
Recherche Astrophysique de Lyon UMR5574, F-69230 Saint-Genis-Laval,
France}
\altaffiltext{5}{ESA/Space Telescope Science Institute (STScI), 3700 San
Martin Drive, Baltimore, MD 21218, USA}
\altaffiltext{6}{Departamento de Astrof{\'i}sica, Facultad de CC.
F{\'i}sicas, Universidad Complutense de Madrid, E-28040 Madrid, Spain}
\altaffiltext{7}{Centro de Astrobiolog\'{\i}a (CAB, INTA-CSIC), Carretera de
Ajalvir km 4, E-28850 Torrej\'on de Ardoz, Madrid, Spain}
\altaffiltext{8}{Universit{\'e} de Toulouse, UPS-OMP, CNRS, IRAP, 9 Av.
colonel Roche, BP 44346, F-31028 Toulouse Cedex 4, France}
\altaffiltext{9}{Observatoire de Gen{\`e}ve, Universit{\'e} de Gen{\`e}ve,
51 Ch. des Maillettes, CH-1290 Sauverny, Switzerland}
\altaffiltext{10}{Department of Astrophysical Sciences, Princeton University,
4 Ivy Lane, Princeton, NJ 08544, USA}

\begin{abstract}
Using the {\it Spitzer Space Telescope} and {\it Herschel Space
Observatory}, we have conducted a survey of infrared galaxies in the field
of the galaxy cluster Abell~S1063 (AS1063) at $z=0.347$, which is one of
the most massive clusters known and a target of the \hst\ CLASH and
Frontier-Field surveys.  The \spitzer/MIPS 24 $\mu$m and \herschel/PACS \&
SPIRE images revealed that the core of AS1063 is surprisingly devoid of
infrared sources, showing only a few detectable sources within the central
r$\sim$1\arcmin.  There is, however, one particularly bright source (2.3
mJy at 24 $\mu$m; 106 mJy at 160 $\mu$m), which corresponds to a background
galaxy at $z=0.61$.  The modest magnification factor (4.0$\times$) implies
that this galaxy is intrinsically IR-luminous (L$_{\rm
IR}=3.1\times10^{11}\ \rm L_{\sun}$).  What is particularly interesting
about this galaxy is that \hst\ optical/near-infrared images show a
remarkably bright and large (1 kpc) clump at one edge of the disk.  Our
follow-up optical/near-infrared spectroscopy shows Balmer (H$\alpha$-H8)
and forbidden emission from this clump (\oii\ $\lambda$3727, \oiii\
$\lambda\lambda$4959,5007, \nii\ $\lambda\lambda$6548,6583), indicating
that it is a \hii\ region.  The \hii\ region appears to have formed in-situ, as
kinematically it is part of a rotating disk, and there is no evidence of
nearby interacting galaxies.  With an extinction correction of A$_{\rm
V}=1.5$ mag, the star formation rate of this giant \hii\ region is $\sim$10
M$_{\sun}$ yr$^{-1}$, which is exceptionally large, even for high redshift
\hii\ regions.  Such a large and luminous \hii\ region is often seen at
$z\sim2$ but quite rare in the nearby Universe.
\end{abstract}


\maketitle

\section{Introduction}
Star forming galaxies at $z\sim2$ have clumpy morphologies \citep{Cowie1995,
Elmegreen2004a, Elmegreen2004b, ForsterSchreiber2011a, ForsterSchreiber2011b}
and form stars more vigorously than seen in the local Universe.  This vigorous
mode of star formation is found in the form of ``clumpy" giant kiloparsec
sized H {\sc ii} regions.  Star formation for galaxies at these redshifts
occurs throughout their disks at higher rates than their local analogs.  There
is evidence that in the local Universe that these clumps form from the
interaction of galaxies.  A canonical picture of this interaction can be seen
in the Antennae galaxies, NGC 4038/39 (Arp 244), where star clusters
form where the material (gas and dust) from both galaxies collide 
\citep{Schweizer1987, Holtzman1992, Whitmore1993, BarnesHernquist1991,
MihosHernquist1996}. However, at high redshift ($z>2$), clumps like these
form in isolation, which is thought to be due to them being gravitationally unstable
and quickly collapsing \citep{Bournaud2007, Dekel2009, Genzel2011}.  

Currently there are two ideas about the formation of these large luminous
clumps; (1) they have same physical process in which they are created at
both low redshift and high redshift, and they are just scaled up \hii\
regions, in which their luminosity scale with with radius, velocity
dispersion and M$_{\rm Jeans}$ \citep{Wisnioski2012}, or (2) the high
redshift clumps are undergoing a different mode star formation, in which
global galaxy properties, such as gas fraction, give rise to the formation
of larger, more luminous clumps suggesting that they evolve with redshift
\citep{Livermore2012, Livermore2015}.  It might be expected that the
physical properties of the clumps evolve with redshift, since the global
star formation rate of the Universe has decreased since the peak at
$z\sim2$ \citep{Madau1998}.  There is some evidence of this seen in the
local Universe in which large luminous star forming regions are quite rare.
Galaxies in the early Universe were more gas rich, leading to greater star
formation; as cool dense gas becomes less available, it might be expected
that the star formation would decrease along with the size of these
regions.  \citet{Guo2015} found in the redshift range of $0.5<z<2$ that
galaxies' clump fraction has decreased over time for higher stellar mass
galaxies, while lower stellar mass galaxies' clump fraction remains almost
constant.  In addition, it is suggested that clumps are short lived; either
clumps migrate toward the bulge of a galaxy or diffuse within 0.1--1 Gyr
\citep{Dekel2009, Genzel2011, Guo2012}.  Conditions at higher redshift,
such as higher gas fractions and cold flow accretion, enabled the regular
formation of these clumps.

Field surveys of galaxies at redshift $z\sim2$ with spatially resolved
H {\sc ii} regions only probe the largest star forming regions, typically
kiloparsecs in size \citep{ForsterSchreiber2011a, ForsterSchreiber2011b}.
  However, this may not necessarily be a reflection of
 ``normal" star forming regions.  There may be many more star forming regions
unresolved to field surveys, which may better reflect the distribution of
normal star forming regions.  In addition, with the evolution of clump
size and luminosity decreasing through time, it makes it more difficult
to detect and characterize intermediate redshift clumps, even with \hst\
resolution.  In the cases where large clumps are detected, it is unclear
whether they truly are a large clump or some combination of smaller
unresolved clumps.

With gravitational lensing it is possible to probe sub-kpc scales of high
redshift galaxies, taking advantage of the image of the galaxy being
magnified and stretched, resolving galaxies at much higher spatial
resolution and enabling the detailed study of star forming regions
\citep{Jones2010a, Livermore2012, Frye2012, Wuyts2014}.  Gravitational
lensing also enables the detection of lower luminosity clumps, where the
flux from relatively faint galaxies is amplified by the lens.  With these
two characteristics, it is possible to detect fainter clumps at sub-kpc
scales, even deblending larger clumps into several smaller clumps 
\citep{Johnson2017a, Johnson2017b, Rigby2017, Cava2018}.

We have seen in the local Universe that star clusters found in star
forming galaxies are bright in the ultra-violet (UV) \citep{Meurer1995}
and far-infrared (far-IR) \citep{Armus1990}.  The far-IR emission is
the re-radiation of UV light from young stars that is scattered and
absorbed by dust.  Star forming regions at high redshift have been
primarily studied in the rest-frame UV due to the availability of high
resolution instrumentation in the optical and near-infrared (near-IR).
Far-IR observations are challenging due to the sensitivity, resolution
and access to currently available instrumentation and facilities.  Only
a limited number of galaxies with clumps at high redshift have been
studied in the far-IR.  In \citet{Wisnioski2013} they investigated the
dust properties of 13 UV selected galaxies at $z=1.3$, from the WiggleZ
sample, and found that only 3 were detected with \herschel.  The
remaining 10 were non-detections, below the sensitivity of \herschel.
Currently with ALMA, \citet{Hodge2016} only could detect the dusty
disks of galaxies at $z\sim2.5$, probing down to kpc scales.  Even
utilizing the longest baselines of ALMA will only have comparable
resolution to \hst, which is not enough to probe sub-kpc scales without
the aide of gravitational lensing.

In order to overcome the difficulty of detecting individual star-forming
clumps at high redshift in the far-IR/submillimeter we use a
gravitationally lensed sample selected with \herschel.  The \herschel\
Lensing Survey \citep[HLS;][]{Egami2010} is a survey of massive galaxy
clusters in the far-IR/submillimeter using \herschel\ to detect
gravitationally lensed galaxies in the submillimeter.  HLS consists of two
surveys, a deep survey (HLS-deep; 290 hrs) of 54 clusters utilizing PACS
(100, 160 $\mu$m) and SPIRE (250, 350, 500 $\mu$m) and a snapshot survey
(HLS-snapshot; 52 hrs) of 527 clusters with SPIRE-only bands.  One of the
main goals of HLS is to identify and follow-up bright
gravitationally-lensed galaxies near the centers of clusters.  It is
expected that there are very few star-forming and post-starburst galaxies
that are cluster members near the projected center of the cluster.  Cluster
cores at low redshift are dominated by passive galaxies. The main
assumption is that the majority of sources emitting in the far-IR near the
cluster center are being gravitationally lensed.  It would be rare to find
cluster galaxies emitting in the far-IR near the cluster core with a
notable exception of brightest cluster galaxies (BCGs) in cool-core
clusters \citep{Rawle2012} and merging clusters, such as Abell
2744 \citep{Rawle2014}.

Abell S1063 (AS1063, RXJ2248-4431) is a particularly interesting cluster in
our sample because it is one of the brightest and most massive galaxy
clusters known with an X-ray luminosity of L$_{\rm X} = (43.2\pm0.6) \times
10^{44}$ erg s$^{-1}$ \citep{Maughan2008} and mass M$_{200} = 33.1^{+9.6}_{-6.8}
\times 10^{14}\ \rm M_{\sun}$ \citep{Gruen2013, Gomez2012, Williamson2011}.  The
X-ray emission \citep{Maughan2008} and Sunyaev-Zel'dovich signal
\citep{Williamson2011} imply that AS1063 is a relaxed cluster, or
virialized.  However, dynamical cluster modelling \citep{Gomez2012} has
shown that it has undergone a recent merger, which is further supported by
the weak lensing analysis by \citet{Gruen2013}.  AS1063's mass and recent
merger history make it an exciting candidate for discovering strongly
lensed high redshift galaxies, such as a quadruply imaged galaxy at
redshift $z=6.1$, \citep{Boone2013, Balestra2013, Monna2014}, and studying
their nebular emission \citep{Mainali2017}.  Massive clusters with ongoing merging, often exhibit large expanded critical lines, increasing the area in which lensed galaxies can be discovered. 
\citet{Gruen2013} has also shown evidence for a possible background cluster
at $z\sim0.6$, which has the potential to be an optimal lensing
configuration for finding highly magnified galaxies due to the chance
alignment of two mass concentrations along the line of sight
\citep{Wong2012}.  AS1063 is one of six Hubble
Frontier Field (HFF) clusters, in which deep \hst\ ACS and WFC3 imaging has
recently been completed with the goal of finding the highest redshift
galaxies and characterizing the populations of galaxies at redshifts
$z=5-10$.   This type of study is beneficial for \herschel-detected
galaxies, as their dusty nature obscures their UV/optical emission, and
greater UV/optical depth in necessarily to detect them.
Within all six HFF clusters, HLS finds $\sim$260 \herschel-detected
galaxies with an optical/near-IR counterpart \citep{Rawle2016}.

In this paper, we report the discovery of three infrared-bright sources in
the core of AS1063. Particularly interesting is the discovery of a luminous
kpc-sized star-forming region in one of these sources, which is a
cluster-lensed infrared luminous galaxy at z=0.6 (AS1063a).  This star
forming region, showing up prominently in the \hst\ optical/near-IR images,
is similar in size and luminosity to clumpy star forming regions found at
higher redshift (z$\sim$2), making this galaxy an excellent lower-redshift
laboratory for studying giant star forming regions at z$\sim$2.  In
addition, the high spatial resolution resulting from lensing magnification
allows us to study in detail the nebular emission properties of this galaxy
and how they relate to the dust and gas.

The paper outline is as follows. In section 2 we present the sample, 
observations, and data reduction methods. In section 3 we present the results
and measurements of the data. In section 4 we discuss the lens model, the physical properties of the galaxy and the giant luminous star forming region
and in section 5 we summarize our results.

The cosmology used throughout this paper is H$_{0}$ = 70 km s$^{-1}$
Mpc$^{-1}$, $\Omega_{\rm M}$ = 0.3, and $\Omega_{\Lambda}$ = 0.7.  All
magnitudes are in AB magnitudes and all flux densities are in mJy.

\section{Observations and Data Reduction}

\subsection{Sample}
The lensed galaxy in this paper comes from HLS-deep \citep{Egami2010}, a
sample of 54 massive galaxy clusters imaged with the \herschel\ Space
Observatory \citep{Pilbratt2010} using PACS (100, 160 $\micron$) and SPIRE
(250, 350 and 500 $\micron$).  The galaxy clusters are selected by X-ray
luminosity, which is a proxy for mass.  With these massive galaxy clusters
it is possible to take advantage of their lensing power to detect faint
high redshift galaxies \citep{Rex2010,Combes2012}.

\subsection{\spitzer\ Imaging and Spectroscopy}
Imaging for AS1063 was obtained (PI: Rieke) at 3.6, 4.5, 5.8, 8.0 $\micron$
using the Infrared Array Camera \citep[IRAC;][]{Fazio2004} on the {\it
Spitzer Space Telescope} \citep{Werner2004}.  Each channel on IRAC has a
field of view (FOV) of 5.2\arcmin$\times$5.2\arcmin\ and pixel size of
$\sim$1.2\arcsec\ pixel$^{-1}$.  The images had a small dither pattern and were
mosaicked together to have a final pixel scale of 0.6\arcsec\ pixel$^{-1}$.  The
integration time was 2400 seconds.  Additional imaging for AS1063 was also
obtained during the warm mission, which is part of the IRAC Lensing Survey
(PI: Egami) adding to the depth of channels 1 and 2 (3.6 and 4.5 $\micron$)
with an integration time of 18000 seconds.  The total integration time for
channels 1 and 2 with the warm and cold missions combined are 20400 seconds.

AS1063 is observed with Multiband Imaging Photometer for \spitzer\
\citep[MIPS;][]{Rieke2004} at 24 $\micron$.  MIPS has a FOV of
5\arcmin$\times$5\arcmin\ and a pixel scale of 2.45\arcsec\ pixel$^{-1}$. The
observations were part of a program to image the fields of clusters in the
Mid-IR (PI: Rieke), the greatest depth covering 6\arcmin$\times$6\arcmin\
of the cluster center with a total integration time of 3600 seconds.  Three
bright sources were identified immediately near the cluster core
(Figure~\ref{fig:ir_images}, top right panel) and were targeted for follow-up.
Photometry for both IRAC and MIPS were measured using {\sc SExtractor}
\citep{BertinArnouts1996} using the parameters \verb'FLUX_AUTO' and
\verb'FLUXERR_AUTO'.

An InfraRed Spectrograph \citep[IRS;][]{Houck2004} spectrum was taken for
the brightest \spitzer/MIPS 24 $\micron$ source, AS1063a (2.5 mJy), in the
cluster core in Long-Low mode (14 $\micron$ - 40 $\micron$) with an
integration of 2400 seconds. IRS in Long-Low mode has a resolution of
R=57--126 and pixel scale of 5.1\arcsec\ pixel$^{-1}$ along the slit.  The 1st and
2nd orders of Long-Low mode have slit widths of 10.7\arcsec\ and
10.5\arcsec\ respectively. Line fluxes in the IRS spectrum were measured
using {\sc pahfit} \citep{JDSmith2007}.

\subsection{\hst\ Imaging}

We used the publicly available \hst\ imaging of AS1063 from the Cluster
Lensing And Supernova survey with Hubble \citep[CLASH;][]{Postman2012}).
CLASH is a survey to image massive galaxy clusters in 16 bands using the
Advanced Camera for Surveys (ACS) and Wide Field Camera 3 (WFC3) with FOVs
of 4\arcmin$\times$4\arcmin\ and 2\arcmin$\times$2\arcmin, on the {\it
Hubble Space Telescope} (\hst).  The CLASH images were released in two
resolutions; 30 mas and 65 mas.  We used the 65mas resolution, for the
increased S/N, especially in the WFC3 where the 30mas images slightly
oversample the PSF.  Photometry was measured using {\sc SExtractor} using
the parameters \verb'FLUX_AUTO'.  Deblending parameters had to be carefully
considered since initial photometry would be able to separate the clump
from the galaxy, but deblending was necessary to remove the nearby
foreground galaxy.  This was also apparent from the publicly released CLASH
catalogs of AS1063.  We used the following {\sc SExtractor} parameters;
\verb'DEBLEND_NTHRESH = 4' and \verb'DEBLEND_MINCONT = 0.005'.  

\subsection{Optical Spectroscopy}

\label{sec:optical_spec}

Optical spectroscopy of AS1063 was obtained with the Low Dispersion Survey
Spectrograph (LDSS-3) on the Magellan-Clay telescope on April 16, 2007 and
July 15-16, 2007.  LDSS-3 is an optical imager/spectrograph with a
4064$\times$4064 STA0500A CCD. LDSS-3 was observed in multi-object mode
using the VPH-all grating, covering a wavelength range of 3750-9500$\AA$,
with a resolution of R=860, an 8.3\arcmin\ diameter FOV and a pixel scale
of 0.189 \arcsec\ pixel$^{-1}$ along each slit. \spitzer/MIPS 24 $\micron$ sources
in AS1063 were selected for spectroscopy, including the three bright 24
$\micron$ sources near the cluster center.  Three masks were observed for
the cluster using 1\arcsec\ width slits.  For the mask containing AS1063a,
the slit was aligned with the major axis of the galaxy based on the
archival \hst/WFPC2 imaging before the CLASH dataset.  Each mask was
integrated for 45 minutes with conditions of 0.76\arcsec, 1.09\arcsec,
0.94\arcsec\ seeing (FWHM).


The LDSS-3 observations were reduced using the COSMOS data reduction
package \citep{Dressler2011}.  COSMOS performs flat-fielding, wavelength
calibration and sky-subtraction which produces a 2D spectrum for each
object. The COSMOS data reduction package is based on an optical model of
model in order to construct a wavelength solution, y-distortion, line
curvature, and tilt for each slit.  The sky-subtraction algorithm is based
on the \citet{Kelson2003} optimal sky-subtraction, in which the
sky-subtraction is performed before rectifying the spectra.  This method
enables better removal of the sky lines and reduces the noise in the final
spectra.  Stacking of the final spectra is based on the positions of the
alignment stars used in the mask. This ensures the maximal amount of flux
per object, accounting for any possible movement in the instrument.  The
dome flats did not have adequate flux in the blue for correcting the
slit-to-slit variation of the sensitivity function introduced by the VPH
grating.  It was necessary to use twilight sky flats to correct for the
slit-to-slit variation.  The 1D spectra were extracted with a 1.9\arcsec\
aperture.  For AS1063a, 5.7\arcsec\ aperture was used for the entire galaxy
with smaller apertures for the individual regions of the galaxy.  Then, the
line fluxes were measured by fitting a Gaussian to the emission lines and
integrating the flux under the curve.  The slit loss for the line flux is
computed by convolving the \hst\ ACS images at F606W and F814W by the
LDSS-3 seeing and measuring the amount flux lost going through the
1\arcsec$\times$12\arcsec\ slit.  Direct images of AS1063 were taken
through mask without the disperser, which were used to determine the
position of the objects within the slits.

\subsection{Near-IR Spectroscopy}

\label{sec:nir_spec}

Near-IR spectroscopy of AS1063a was obtained with the MMT and Magellan
Infrared Spectrograph \citep[MMIRS;][]{McLeod2012} on the Magellan-Clay
telescope in longslit mode on April 14, 2012, under poor seeing
($>$1.66\arcsec) and non-photometric conditions. MMIRS is a near-IR
imager/spectrograph with 2048$\times$2048 HgCdTe Hawaii-2 detector. The
spectra were taken with a 1.2\arcsec\ width slit using the J grating with
the zJ filter, covering a wavelength range of 0.94-1.51 $\micron$, with a
resolution of R=800, a 6.9\arcmin$\times$6.9\arcmin\ FOV and pixel scale of
0.202 \arcsec\ pixel$^{-1}$ along the slit.  The spectra were taken in 5 minute
exposures using a 4 position dither pattern for 4 exposures providing a
total integration of 20 minutes on source.  The longslit alignment was with
the major axis of AS1063a.  The spectra were flat-fielded with dome flats,
and wavelength calibrated by OH skylines. Frames were subtracted from each
other (A-B, B-A) in order to remove the sky emission from the source as
well as the dark current.  All the frames were median combined into one
image using \verb'imcombine' in {\sc iraf} \citep{Tody1986,Tody1993},
with sigma clipping turned on.  The 1D spectra were extracted with a
8\arcsec\ aperture.  The line fluxes were measured by fitting a
Gaussian to the emission lines and integrating the curve.  Then they
were corrected for slit loss by comparing the ratio of the \hst\ images
at F105W to the same image convolved by the MMIRS seeing.  The slit
loss for the line flux is computed by convolving the \hst\ WFC3/IR
images at F105W by the MMIRS seeing and measuring the amount flux lost
going through the slit.
 
\hst\ WFC3/IR G102 and G141 grism spectra for AS1063 were obtained from the
Grism Lens-Amplified Survey from Space \citep[GLASS;][]{Schmidt2014,
Treu2015}. The grism data for AS1063 was reduced using {\sc aXe}
\citep{Kummel2009}, which maps objects detected in the direct image to the
objects in slitless grism spectra.  The direct images were combined using
MultiDrizzle.  Sources in the direct image were detected using SExtractor,
which is used as an input catalog for {\sc aXe}.  A contamination model is
created from all the detected sources, including their 0th, 1st and 2nd
order spectra.  Tweakshifts is then used to determine their spatial offset
between direct images for different visits but at the same roll angle.
Then, the grism spectra of multiple visits, at the same roll angle, were
drizzled together.  Finally, the {\sc aXe} routines are run to drizzle the
2D spectra and extract the spectra with the options of
\verb'slitless_geom' and
\verb'orient' turned on.  The 2D spectra are then divided by the instrument
sensitivity function to flux calibrate the spectra.  A sliding median with
a window of 50 pixels (1200 \AA) was used to subtract off the continuum and
contaminating sources from AS1063a.  Apertures were placed over the
emission lines in the 2D spectra to measure their fluxes. 

\subsection{Far-IR/Submillimeter Imaging}

\herschel\ PACS \citep{Poglitsch2010} images at 100 and 160 $\micron$ were
obtained for the core of AS1063 with a FOV of 9\arcmin$\times$9\arcmin\
\citep{Egami2010}.  The PACS instrument, operating in the dual band
photometry mode, consists of two bolometer arrays; a blue channel
(32$\times$64 pixels) and a red channel (16$\times$32 pixels).  The blue
channel has two filters, 60-85 $\micron$ and 85-125 $\micron$ whereas the
red channel has one, 125 - 210 $\micron$, with each channel having a FOV of
1.75\arcmin$\times$3.5\arcmin.  Additional 70 $\micron$ images were later
obtained (PI: T. Rawle) in May 26, 2013, to better constrain the dust
temperatures of warm cluster galaxies.  Since PACS operates in a dual
photometer mode, deeper 160 $\micron$ images were also obtained with this
program.  The PACS maps were generated with {\sc UniMap}
\citep{Piazzo2015} with a pixel scale
of 1.0\arcsec, 1.0\arcsec\ and 2.0\arcsec.

\herschel\ SPIRE \citep{Griffin2010} images at 250, 350 and 500 $\micron$ were
obtained for the field of AS1063 with a FOV of 23\arcmin$\times$26\arcmin.
SPIRE in photometry mode has three bands, 250, 350 and 500 $\micron$, with
the bands consisting of 139, 88 and 43 bolometers respectively.  Each band
has a FOV of 4\arcmin$\times$8\arcmin.  The FWHM of the beam size at each
band is 17.6\arcsec, 23.9\arcsec\ and 35.2\arcsec. SPIRE maps were
generated using HIPE v10.0.  The pixel scales of each of the maps are
6\arcsec, 9\arcsec\ and 12\arcsec, respectively.   SPIRE photometry was
measured using {\sc iraf} task \verb'daophot'. 

Large APEX Bolometer Camera \citep[LABOCA;][]{Siringo2009} 870 \micron\
observations of AS1063 were part of the LABOCA Lensing Survey (E187A0437A,
M-087.F-0005-2011).  The LABOCA 870 \micron\ images of AS1063 have FOV is
$\sim$5.4\arcmin, with the center of the field 1\arcmin\ north of the
cluster center. The LABOCA beam is 24.3\arcsec\ (FWHM).  More details about
the observation, reduction photometry can be found in \citet{Boone2013}.  

\section{Results}

\subsection{Three Bright 24 $\mu$m Sources}

\begin{figure*}
\begin{center}
\includegraphics[scale=1.0]{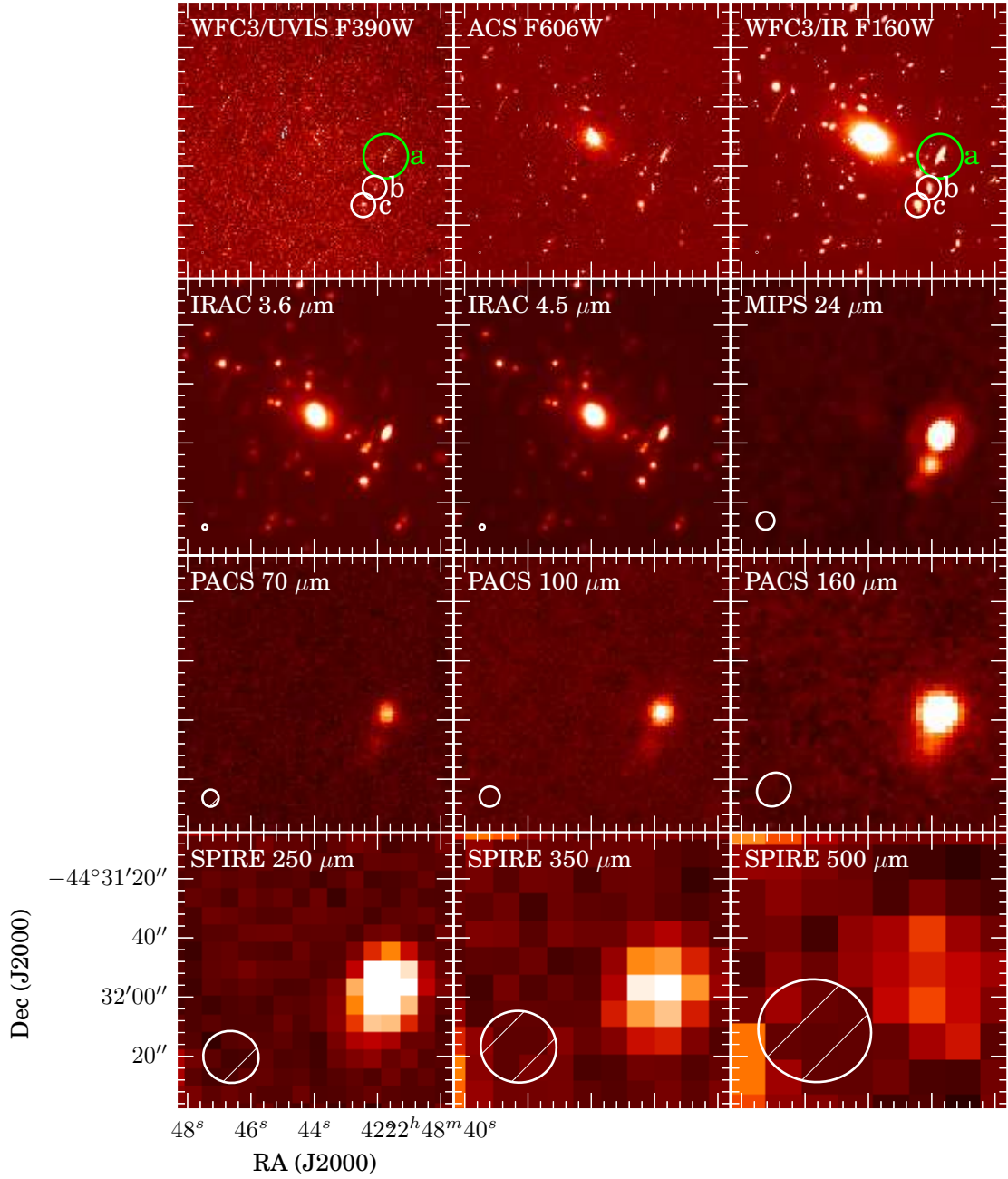}
\caption{\label{fig:ir_images}
The cluster core of AS1063 imaged in \hst\ (WFC3/UVIS F390W, ACS F606W,
WFC3/IR F160W), \spitzer\ IRAC (3.6 and 4.5 $\micron$) and MIPS 24
$\micron$, \herschel\ PACS (100 and 160 $\micron$) and SPIRE (250, 350 and
500 $\micron$). Each panel shows the central 1.6\arcmin$\times$1.6\arcmin\
area of the cluster core. The green circle marks the position of the lensed galaxy AS1063a.  The lensed galaxy's far-IR/submm SED is fully sampled by PACS and SPIRE. The white circles mark the positions of the two cluster galaxies (z$_{\rm spec}$ = 0.336) which can be seen in MIPS and PACS, but drop out of the SPIRE bands. The PSF of each of the instruments is marked in the bottom left corner in each
panel.}

\end{center}
\end{figure*}

Figure~\ref{fig:ir_images} shows the \spitzer/IRAC (3.6 and 4.5 $\mu$m),
\spitzer/MIPS (24 $\mu$m), \herschel/PACS (70, 100, 160 $\mu$m), and
\herschel/SPIRE (250, 350, and 500 $\mu$m) images covering the
central 1.6\arcmin$\times$1.6\arcmin\ area of the massive cluster
AS1063.  As the figure shows, the cluster core is surprisingly
devoid of infrared/submillimeter sources, but three bright sources
are clearly detected 24\arcsec\ southwest of the brightest cluster
galaxy (BCG).  Their IRAC counterparts were unambiguously identified
(marked with the green and white circles in Figure~\ref{fig:ir_images}),
and the brightest 24 $\mu$m source in the north is seen to dominate
the observed fluxes in the \herschel/PACS and SPIRE bands. The
measured flux densities of these three bright 24 $\mu$m sources are
listed in Table~\ref{tab:photometry}. 

\begin{figure*}
\begin{center}
\includegraphics[scale=0.46]{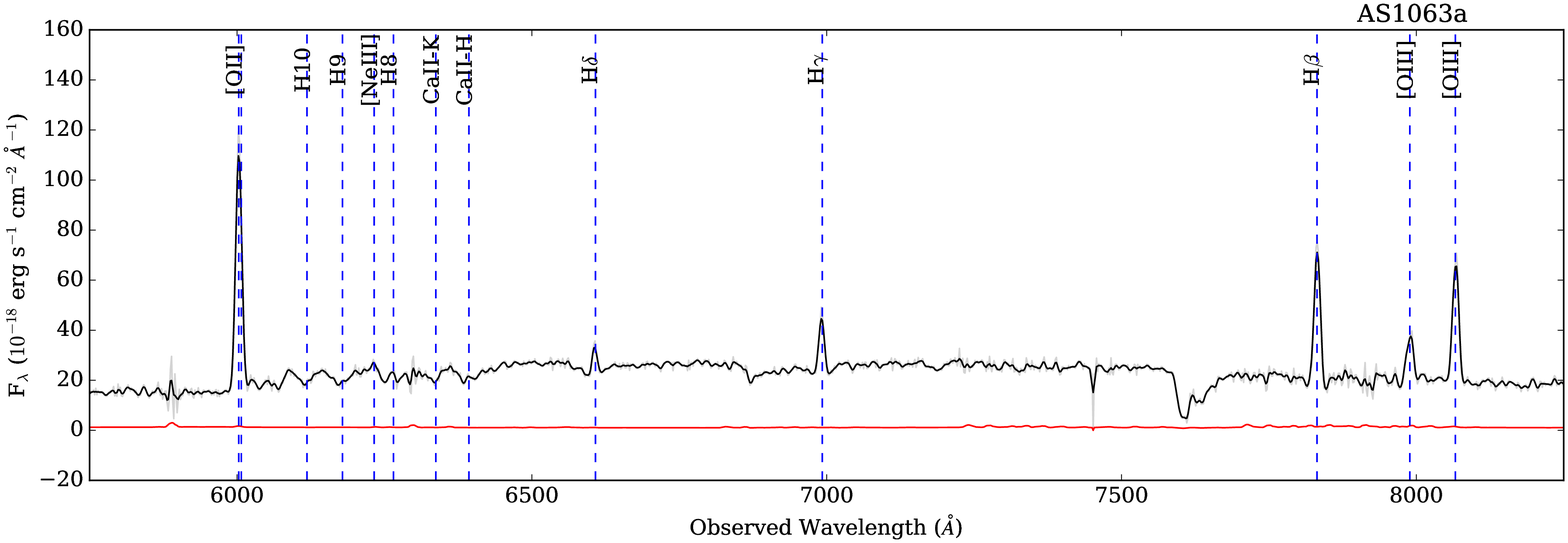}
\includegraphics[scale=0.46]{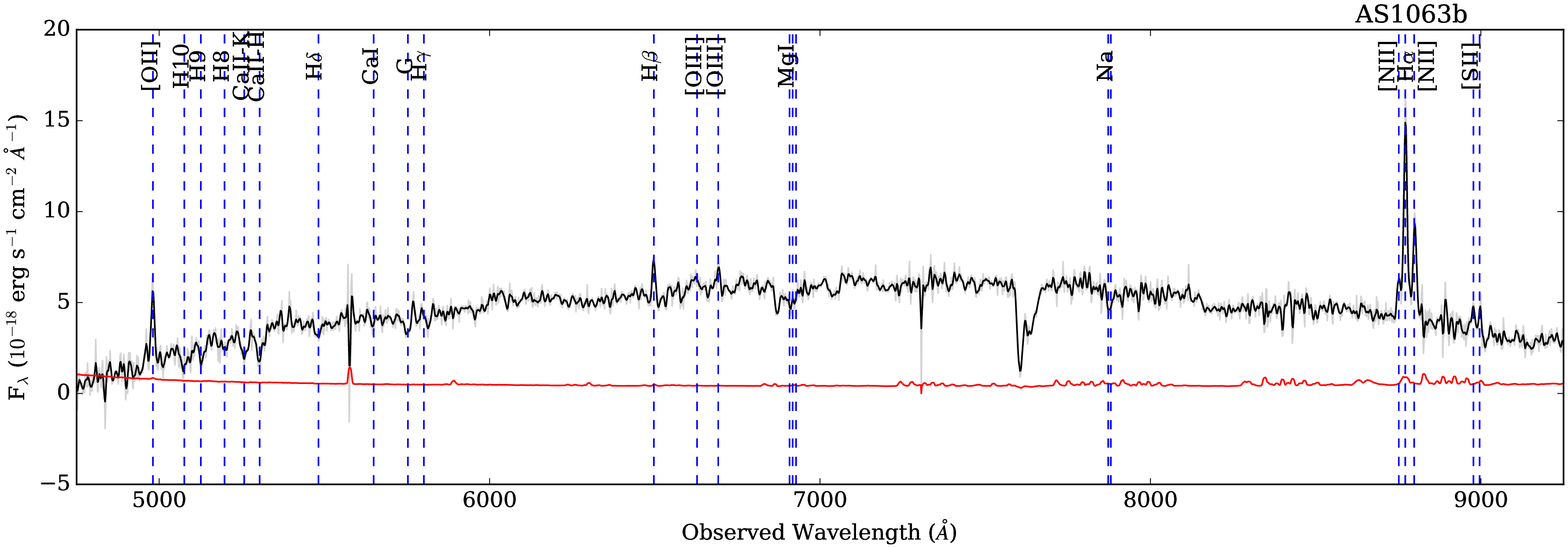}
\includegraphics[scale=0.46]{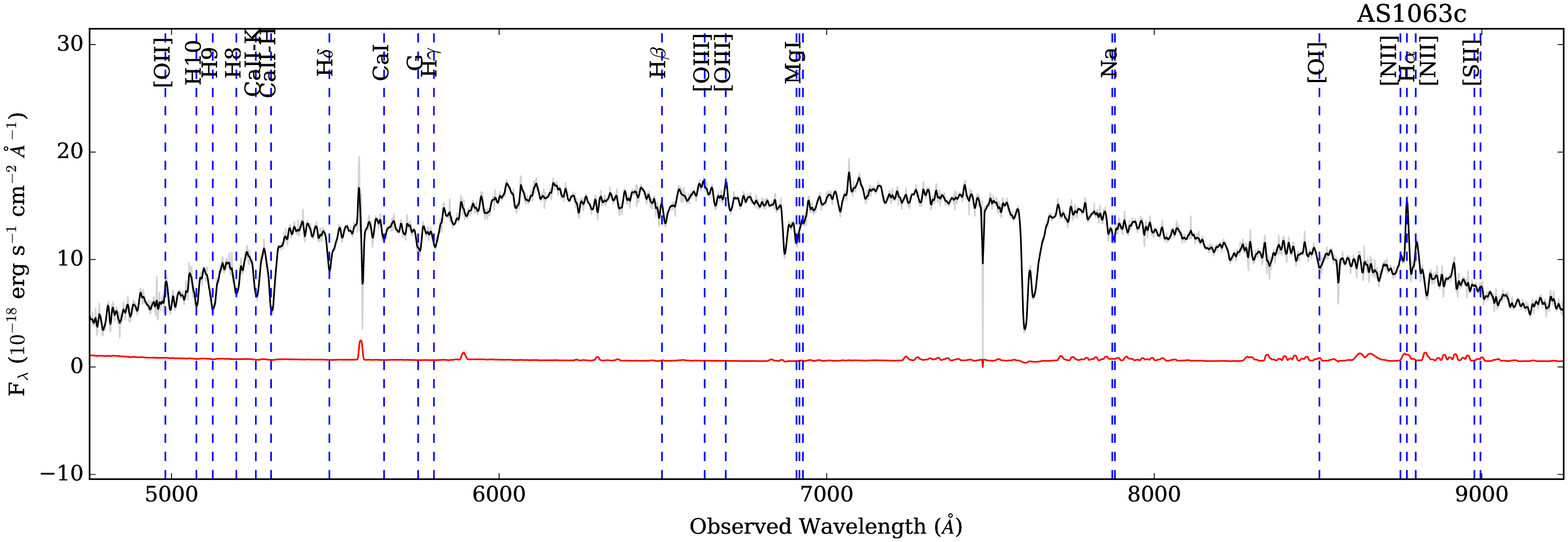}
\end{center}
\caption{\label{fig:ldss3_spectra}Magellan/LDSS-3 optical spectra of the
three bright 24 $\micron$ sources in AS1063. (Top) Multiple Balmer lines
(H$\alpha$ -- H8) are detected in AS1063a suggesting it has undergone a
recent burst of of star formation.  (Middle and Bottom) We detect H$\alpha$
in the cluster members (AS1063b and AS1063c) suggesting that they may be
falling into the cluster for the first time.} 
\end{figure*}

\floattable
\begin{deluxetable}{lccc} 
\tabletypesize{\footnotesize}
\tablewidth{0pc}
\tablecaption{\label{tab:photometry}Photometry of Three Bright 24 $\micron$ Sources}
\tablecolumns{4}
\tablehead{
\colhead{} & \colhead{AS1063a} & \colhead{AS1063b} & \colhead{AS1063c} \\
\hline
\colhead{Band} & \colhead{Magnitude} & \colhead{Magnitude} & \colhead{Magnitude} \\
\colhead{} & \colhead{[AB]} & \colhead{[AB]} & \colhead{[AB]}
}
\startdata
F225W  & 21.48$\pm$0.05 & $<$22.88       & $<$23.41 \\   
F275W  & 21.01$\pm$0.03 & 22.46$\pm$0.15 & 22.24$\pm$0.08 \\
F336W  & 20.96$\pm$0.05 & 22.33$\pm$0.11 & 21.90$\pm$0.05 \\
F390W  & 20.94$\pm$0.07 & 21.83$\pm$0.04 & 21.67$\pm$0.04 \\
F435W  & 20.68$\pm$0.06 & 21.57$\pm$0.16 & 21.37$\pm$0.06 \\
F475W  & 20.55$\pm$0.08 & 21.47$\pm$0.12 & 20.85$\pm$0.07 \\
F606W  & 19.81$\pm$0.04 & 20.50$\pm$0.04 & 19.86$\pm$0.03 \\
F625W  & 19.58$\pm$0.06 & 20.28$\pm$0.06 & 19.62$\pm$0.04 \\
F775W  & 19.10$\pm$0.04 & 19.86$\pm$0.05 & 19.22$\pm$0.03 \\
F814W  & 19.01$\pm$0.05 & 19.68$\pm$0.06 & 19.13$\pm$0.04 \\
F850LP & 18.85$\pm$0.08 & 19.52$\pm$0.09 & 18.95$\pm$0.06 \\
F105W  & 18.54$\pm$0.01 & 19.30$\pm$0.02 & 18.77$\pm$0.01 \\
F110W  & 18.47$\pm$0.01 & 19.16$\pm$0.02 & 18.67$\pm$0.01 \\
F125W  & 18.42$\pm$0.02 & 19.02$\pm$0.03 & 18.54$\pm$0.02 \\
F140W  & 18.25$\pm$0.01 & 18.89$\pm$0.01 & 18.42$\pm$0.01 \\
F160W  & 18.07$\pm$0.02 & 18.75$\pm$0.02 & 18.31$\pm$0.02 \\
\hline
Band & Flux  & Flux  & Flux \\
     & [mJy] & [mJy] & [mJy] \\
\hline
3.6 $\mu$m & 0.35$\pm$0.03 & 0.13$\pm$0.01 & 0.18$\pm$0.02 \\
4.5 $\mu$m & 0.29$\pm$0.03 & 0.12$\pm$0.01 & 0.17$\pm$0.02 \\
5.8 $\mu$m & 0.37$\pm$0.04 & 0.09$\pm$0.01 & 0.10$\pm$0.01 \\
8.0 $\mu$m & 0.30$\pm$0.04 & 0.22$\pm$0.02 & 0.12$\pm$0.01 \\
24 $\mu$m  & 2.22$\pm$0.02 & 0.77$\pm$0.01 & 0.27$\pm$0.03\\
70 $\mu$m  & 32.2$\pm$2.3  & 9.0$\pm$0.5   & $<$3.1 \\
100 $\mu$m & 69.3$\pm$4.9  & 14.5$\pm$1.0  & 3.7$\pm$0.8 \\
160 $\mu$m & 105.8$\pm$7.5 & 24.8$\pm$1.3  & $<$8.2 \\
250 $\mu$m & 69.3$\pm$6.8  & 11.9$\pm$5.4  & $<$16.2 \\
350 $\mu$m & 36.1$\pm$6.2  & $<$17.1       & $<$17.1  \\
500 $\mu$m & 21.9$\pm$6.1  & $<$17.5       & $<$17.5  \\
\enddata
\tablecomments{
Herschel errors listed are computed from the RMS of the maps plus the
calibration error, which is 5\% in PACS and 4\% in SPIRE. PACS and SPIRE
flux limits were presented in \citet{Rawle2016}.
}

\end{deluxetable}

Our follow-up optical spectroscopy showed that the brightest 24 $\mu$m
source (AS1063a) corresponds to a background galaxy at $z=0.61$, detecting
Balmer emission (H$\beta$-H8), \oii $\lambda$3727, \oiii\
$\lambda\lambda$4959,5007, and \neiii).  The other two fainter 24 $\mu$m
sources (AS1063b and AS1063c) correspond to cluster galaxies at $z=0.347$,
detecting Ca {\sc ii} H and K absorption and H$\alpha$ and \nii\
$\lambda\lambda$ 6548,6583.  The three 24 $\micron$ sources are listed in
Table~\ref{tab:bright} and their optical spectra are plotted in
Figure~\ref{fig:ldss3_spectra}).  These redshifts are consistent with those
published by \citet{Gomez2012}. 

\begin{figure}
\begin{center}
\hspace{1.8em}\includegraphics[scale=0.331]{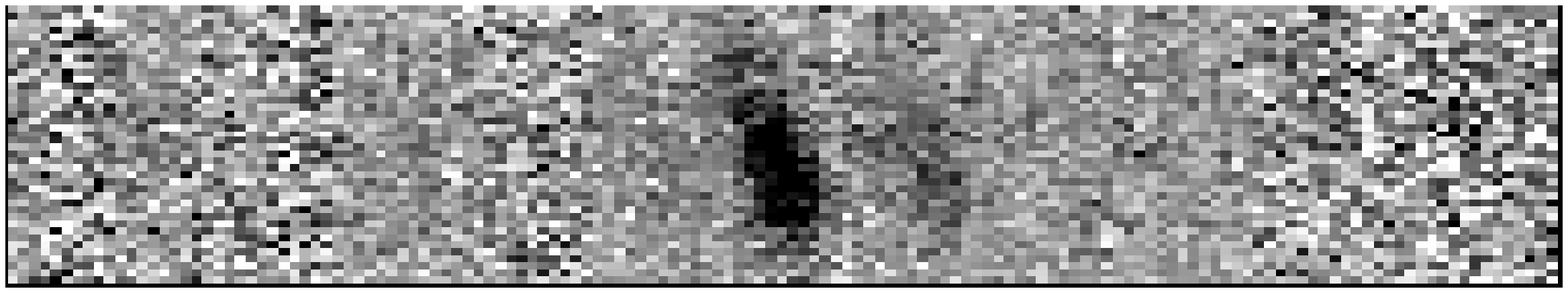}
\includegraphics[scale=0.35]{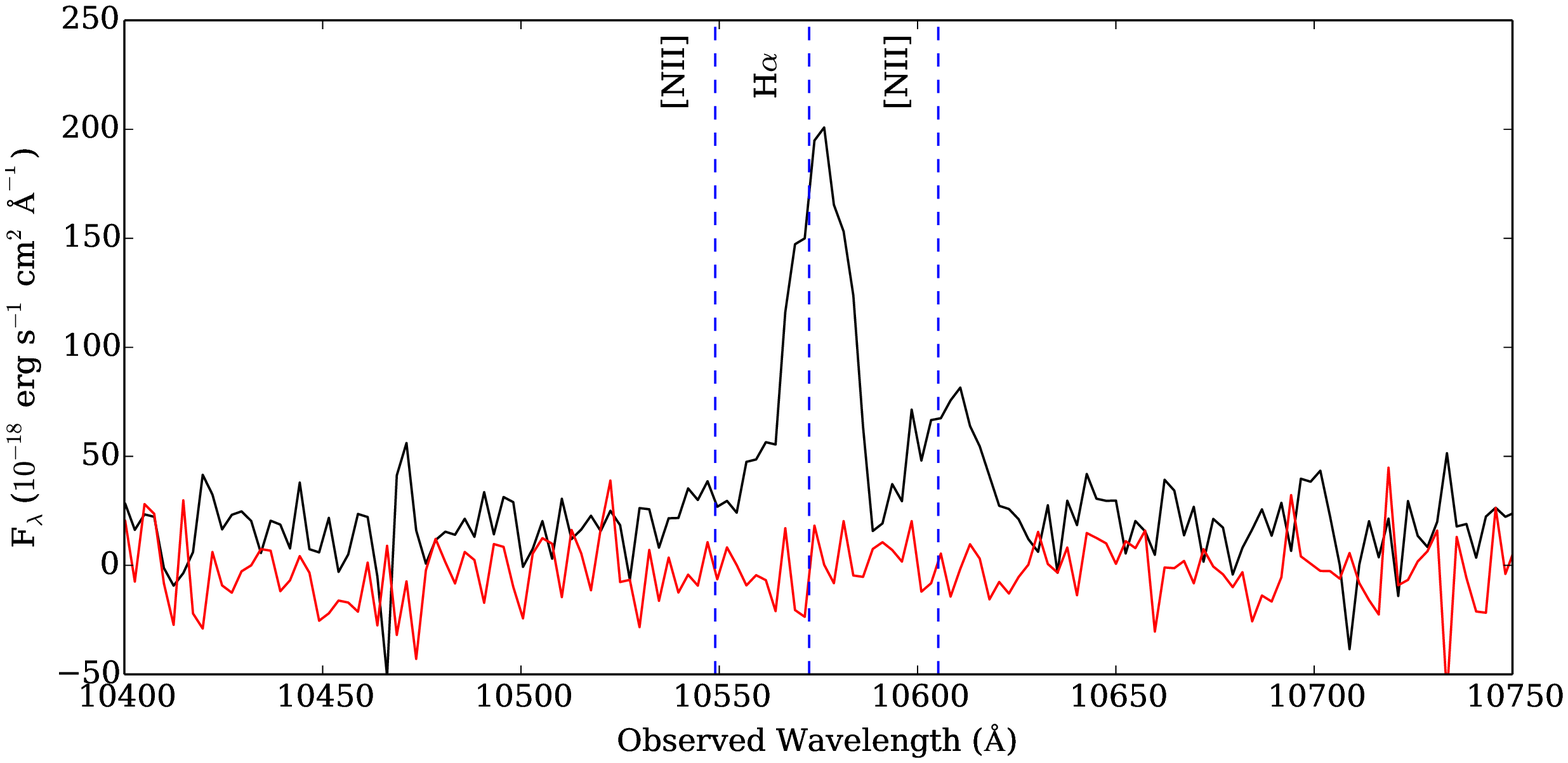}
\end{center}
\caption{\label{fig:mmirs_spectrum}Magellan/MMIRS spectrum of AS1063a,
H$\alpha$ and \nii$\lambda$6585 are clearly detected within 20 minutes of
integration. The asymmetry of the line is the result of the velocity
offset of the bright clump at the edge of the galaxy.  Even though the
magnification is increasing across the galaxy towards the clump (southeast
direction), it is not a major contributor to the asymmetry.}
\end{figure}

\begin{figure}
\begin{center}
\includegraphics[scale=0.45]{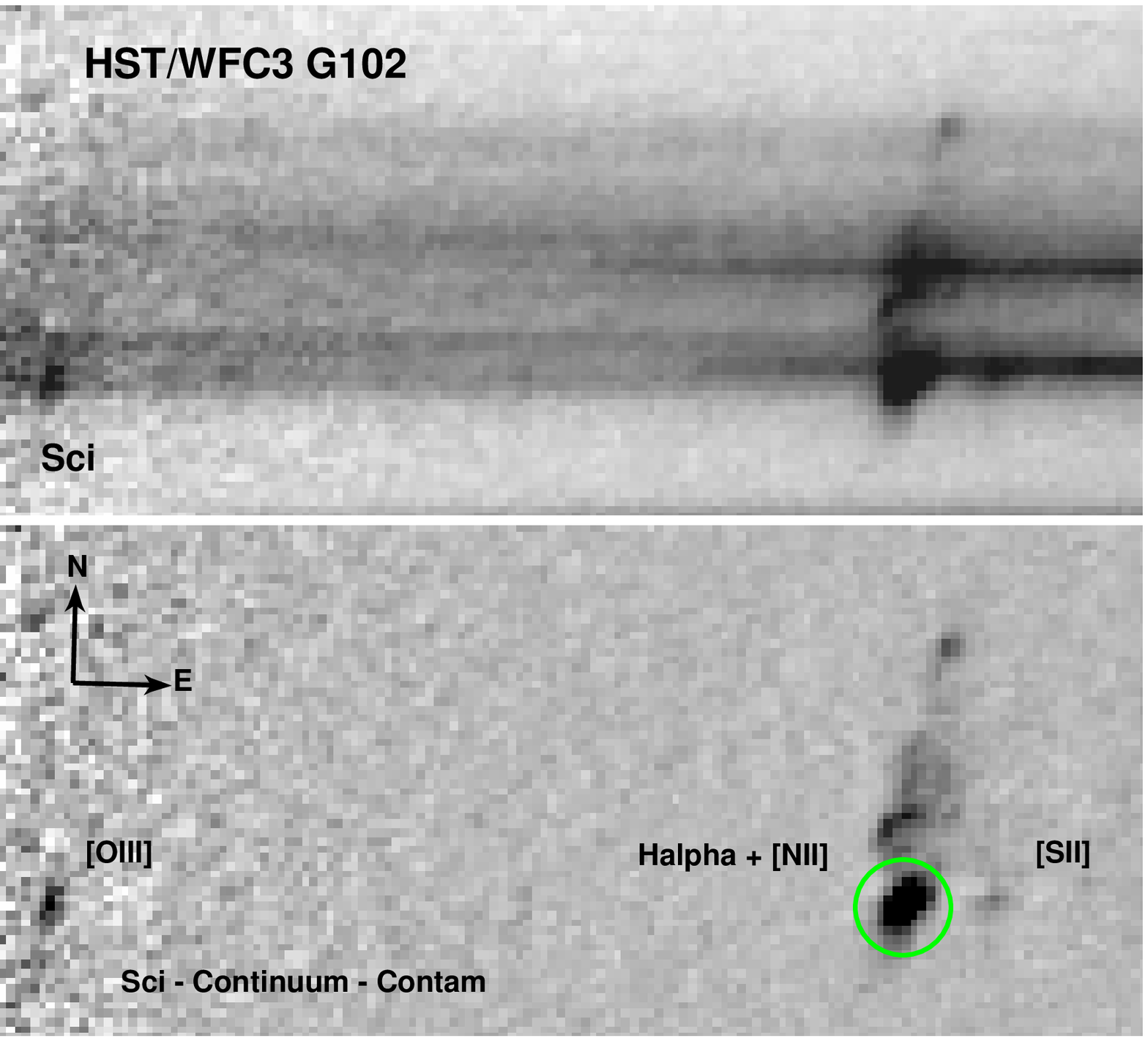}
\end{center}
\caption{\label{fig:grism_spectra}\hst\ WFC3/IR G102 spectrum of the lensed
galaxy AS1063a. (top panel) The reduced {\sc aXe} spectrum. (bottom panel)
The green circle marks the position of the star forming region.  The
reduced {\sc aXe} spectrum with the contaminating sources subtracted and
AS1063a continuum subtracted, using a sliding median.
H$\alpha$+\nii$\lambda\lambda$6548,6583 are detected throughout the entire
galaxy, while \oiii$\lambda$5007 and \sii$\lambda\lambda$6717,6731 are only
detected for the star forming region.}
\end{figure}

The MMIRS near-infrared spectrum of AS1063a detect H$\alpha$ and
\nii$\lambda$6585 lines (Figure~\ref{fig:mmirs_spectrum}) and the \hst\
WFC3/IR G102 grism spectrum detect H$\alpha$ blended with
\nii$\lambda\lambda$6548,6583 lines (Figure~\ref{fig:grism_spectra}). In
addition, the grism also detects \sii$\lambda$6717,6731 and
\siii$\lambda$9069,9545.  The measured line fluxes are shown in
Table~\ref{tab:ldss3_mmirs} and \ref{tab:grism}.  The line fluxes corrected
for slit loss are also given, as described in
Sections~\ref{sec:optical_spec} and \ref{sec:nir_spec}.  The \hst\ WFC3/IR
G102 grism observations of H$\alpha$+\nii\ is consistent with the
Magellan/MMIRS measurement of those lines.
\floattable
\begin{deluxetable}{lrr} 
\tabletypesize{\footnotesize}
\tablewidth{0pc}
\tablecaption{\label{tab:ldss3_mmirs}Optical and Near-Infrared Emission Line Fluxes of AS1063a}

\tablehead{
\colhead{Line} & \colhead{Flux} & \colhead{Flux$_{\rm corr}$$^{a}$} \\
\colhead{} & \colhead{($10^{-17} {\rm erg/s/cm}^{2}$)} &
\colhead{($10^{-17} {\rm erg/s/cm}^{2}$)}
}
\startdata
\multicolumn{3}{c}{Entire galaxy} \\
\hline
\oii$\lambda$3727   & 123.8$\pm$1.5 & 132.1$\pm$1.6 \\
H$\delta$              &  10.3$\pm$0.8 &  11.0$\pm$0.9 \\
H$\gamma$              &  25.9$\pm$1.1 &  27.7$\pm$1.2 \\
H$\beta$               &  74.4$\pm$2.0 &  80.0$\pm$2.1 \\
\oiii$\lambda$4959  &  33.0$\pm$2.8 &  35.3$\pm$3.1 \\
\oiii$\lambda$5007  &  67.4$\pm$1.7 &  72.0$\pm$1.8 \\
H$\alpha$              & 308.1$\pm$8.2 & 348.8$\pm$9.3 \\
\nii$\lambda$6583   & 137.7$\pm$6.9 & 155.9$\pm$7.8 \\ 
\hline
\hline
\multicolumn{3}{c}{H {\sc ii} region} \\
\hline
\oii$\lambda$3727   &  61.6$\pm$0.8 &  65.8$\pm$0.9 \\
\neiii$\lambda$3869 &   2.9$\pm$0.6 &   3.1$\pm$0.6 \\
H8                     &   4.2$\pm$0.5 &   4.5$\pm$0.5 \\
H$\epsilon$            &   2.8$\pm$0.4 &   3.0$\pm$0.4 \\
H$\delta$              &   7.0$\pm$0.4 &   7.5$\pm$0.4 \\
H$\gamma$              &  17.1$\pm$0.5 &  18.4$\pm$0.5 \\
H$\beta$               &  42.3$\pm$0.7 &  45.5$\pm$0.8 \\
\oiii$\lambda$4959  &  15.5$\pm$0.7 &  16.7$\pm$0.8 \\
\oiii$\lambda$5007  &  48.3$\pm$0.6 &  51.9$\pm$0.6 \\
\hline
\multicolumn{3}{c}{Bulge} \\
\hline
\oii$\lambda$3727   &  40.2$\pm$0.9 &  43.0$\pm$1.0 \\
H$\delta$              &   2.9$\pm$0.4 &   3.1$\pm$0.4 \\
H$\gamma$              &   8.4$\pm$0.7 &   9.0$\pm$0.7 \\
H$\beta$               &  25.5$\pm$1.1 &  27.3$\pm$1.2 \\
\oiii$\lambda$4959  &   7.4$\pm$1.4 &   7.9$\pm$1.5 \\
\oiii$\lambda$5007  &  14.1$\pm$0.8 &  15.1$\pm$0.9 \\
\hline
\multicolumn{3}{c}{Spiral arm} \\
\hline
\oii$\lambda$3727   &   9.1$\pm$0.5 &   9.7$\pm$0.5 \\
H$\gamma$              &   1.3$\pm$0.3 &   1.4$\pm$0.3 \\
H$\beta$               &   2.4$\pm$0.6 &   2.6$\pm$0.6 \\
\oiii$\lambda$5007  &   2.4$\pm$0.3 &   2.6$\pm$0.3 \\
\hline
\enddata

\tablecomments{
$^{(a)}$ Flux corrected for slitloss by convolving the \hst\ ACS and WFC/IR
images by the LDSS-3 and MMIRS seeing and measuring the ammount of light
that is blocked by the slit.
}

\end{deluxetable}

\floattable
\begin{deluxetable}{lr} 
\tabletypesize{\footnotesize}
\tablewidth{0pc}
\tablecaption{\label{tab:grism}\hst/WFC3 G102 and G141 Spectra of AS1063a}

\tablehead{
\colhead{Line} & \colhead{Flux} \\
\colhead{} & \colhead{($10^{-17} {\rm erg/s/cm}^{2}$)} 
}
\startdata
\multicolumn{2}{c}{Entire galaxy} \\
\hline
H$\alpha$+\nii$\lambda\lambda$6548,6583  & 545.5$\pm$2.4 \\
\hline
\hline
\multicolumn{2}{c}{H {\sc ii} region} \\
\hline
\oiii$\lambda$5007                       &  53.2$\pm$4.1 \\
H$\alpha$+\nii$\lambda\lambda$6548,6583  & 275.0$\pm$1.2 \\
\sii$\lambda\lambda$6717,6731            &  15.1$\pm$0.6 \\ 
\siii$\lambda$9069                       &   4.4$\pm$0.6 \\ 
\siii$\lambda$9545                       &  32.5$\pm$0.7 \\ 
\hline
\multicolumn{2}{c}{Bulge} \\
\hline
H$\alpha$+\nii$\lambda\lambda$6548,6583  & 205.1$\pm$1.5 \\
\hline
\multicolumn{2}{c}{Spiral arm} \\
\hline
H$\alpha$+\nii$\lambda\lambda$6548,6583  &  50.5$\pm$1.2 \\
\hline
\enddata


\end{deluxetable}

AS1063a is detected in all three PACS bands (Figure \ref{fig:ir_images}).
At 70 and 100 $\micron$ it is distinctly identifiable whereas at 160
$\micron$ it becomes blended with AS1063b. The third bright source (AS1063c)
is detected at 70 $\micron$ and marginally detected at 100 and 160
$\micron$.  The sources were not severely crowded and it was possible to get
similar photometric values ($<7\%$ difference) doing both aperture and
point spread function (PSF) photometry.  Aperture photometry was used at 70
and 100 $\micron$ and PSF photometry was used at 160 $\micron$.

\begin{figure}
\begin{center}
\includegraphics[scale=0.42]{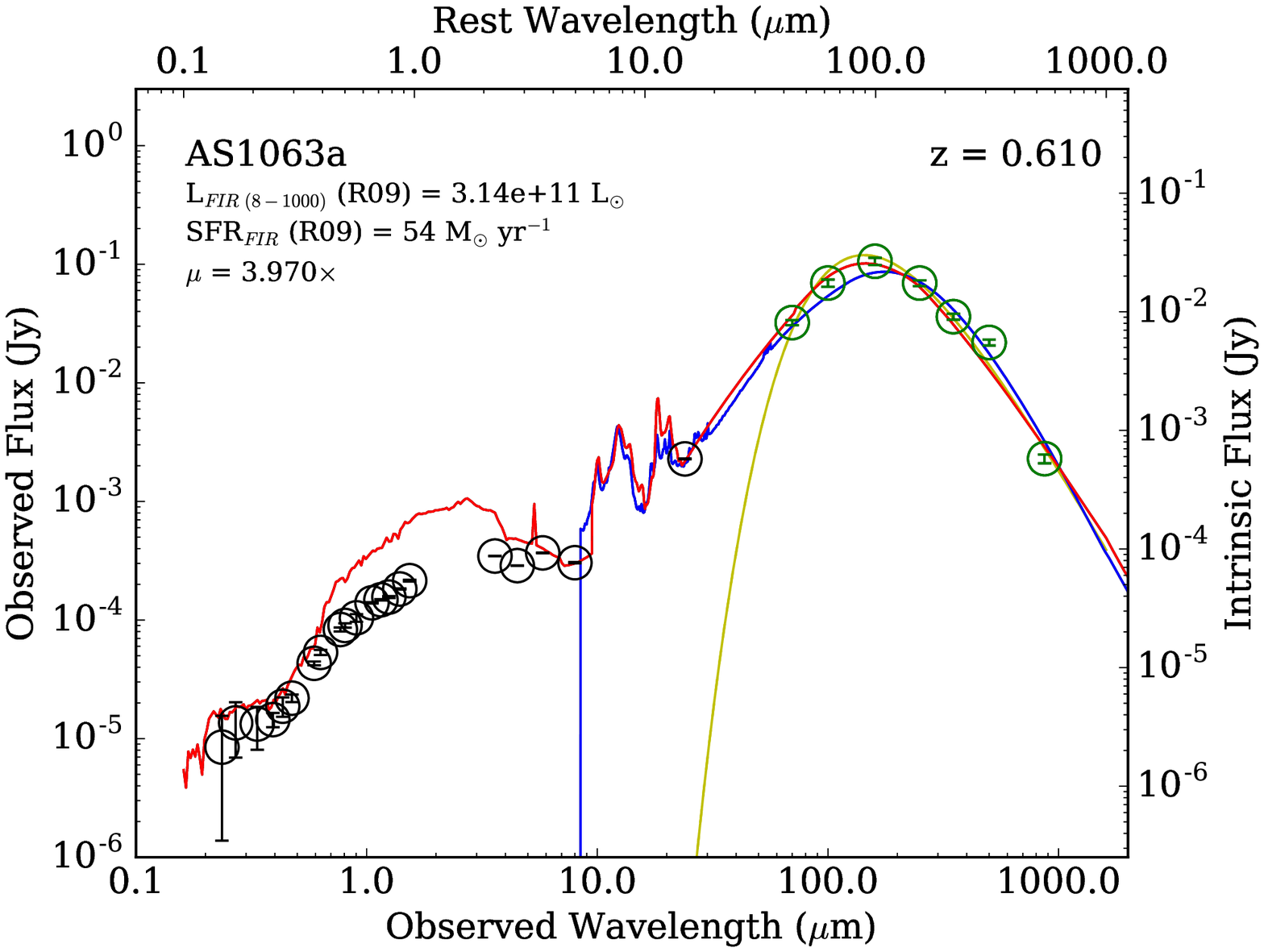}
\caption{\label{fig:SED}
SED of the lensed galaxy AS1063a in the core of AS1063.  Fits were made to
the \herschel\ PACS 70, 100, 160 $\micron$ and \herschel\ SPIRE 250, 350,
500 and LABOCA 870 $\micron$ (green circles).  The photometry blueward of
the 70 $\micron$ was ignored in the fitting of the SED (black circles).  
The lines shows the best fitting templates to the far-IR/submm data. The
blue line is the \citet{Rieke2009} templates and the red line is the
\citet{CharyElbaz2001} templates.  The yellow line shows the best fitting
modified blackbody.
On the right axis, the observed flux was corrected based on the
magnification (4.0$\pm$0.1) determined by the lens modelling by {\sc
lenstool} \citep{Jullo2007}.} 

\end{center}
\end{figure}

At 250 $\micron$ AS1063a was blended with the nearby cluster galaxy
AS1063b, and crowded field photometry (PSF fitting) was necessary to
properly deblend the sources.  The SED for AS1063a is shown in Figure
~\ref{fig:SED}.  Within the cluster field it was difficult to find bright
well-isolated sources to measure the PSF of the image, so we used an
empirical PSF provided by the \herschel\ Science Center.  The empirical PSF
was binned and rotated to the PA of \herschel\ when it observed AS1063 and
after subtracting the sources from the map had a resulting RMS of $\sim$5.9
mJy.  At longer wavelengths (350 and 500 $\micron$) the two cluster members
are almost completely undetected.

In order to determine the optical/near-IR counterpart of a submillimeter
source it is necessary to have several bands spanning a wide wavelength
range between the optical and the submillimeter.  Sources are traced from
the submillimeter to the optical, stepping down in wavelength while
ensuring that each source is being followed near the centroid of the
original submillimeter source.  Sometimes submillimeter sources may be
blends of multiple IRAC and MIPS sources.  If the sources are not too close
(less than a pixel away), then they can be deblended using the {\sc iraf}
task \verb'daophot' with the IRAC and MIPS position priors.  The redshift
of the sources can also alleviate source confusion, where dust emission
from lower redshift sources may fall below the detection limit for longer
wavelength SPIRE bands.

\floattable
\begin{deluxetable}{rrrccc}
\tabletypesize{\footnotesize}
\tablewidth{0pc}
\tablecaption{\label{tab:bright}Spectroscopic Redshifts of Three Bright 24 $\micron$ Sources}
\tablecolumns{6}
\tablehead{
\colhead{Source}  & \colhead{R.A.} & \colhead{Decl.} &
\colhead{z$^{\tablenotemark{a}}$} &
\colhead{z$^{\tablenotemark{b}}_{\rm quality}$} &
\colhead{ID$^{\tablenotemark{c}}$}
}
\startdata
AS1063a & 22:48:41.760 & -44:31:56.53 & 0.611 & 4 & 30 \\
AS1063b & 22:48:42.113 & -44:32:07.39 & 0.337 & 4 & 8  \\
AS1063c & 22:48:42.480 & -44:32:12.99 & 0.337 & 4 & 9  \\
\enddata
\tablenotetext{a}{Spectroscopic redshift originally published in \citet{Gomez2012}}
\tablenotetext{b}{z$_{\rm quality}$ is described in Appendix \ref{sec:append}}
\tablenotetext{c}{ID of galaxy used in Table \ref{tab:24um_src}}

\end{deluxetable}

A complete list of 24 $\mu$m sources with spectroscopic redshifts within
the deepest MIPS coverage of AS1063 (5\arcmin$\times$5\arcmin) is presented
in Appendix \ref{sec:append}.  A full analysis will be presented in a
future paper.

\subsection{IR-Luminous Lensed Galaxy at $z=0.61$}

\label{sec:ir_luminous_galaxy}

In the infrared/submillimeter range, the most conspicuous source in the
core of AS1063 is the infrared-bright galaxy at $z=0.61$
(Figure~\ref{fig:ir_images}).  Its SED was fit using \citet{CharyElbaz2001} and
\citet{Rieke2009} templates (Figure~\ref{fig:SED}).  Both template
sets are based on SEDs of local galaxies.  The best $\chi^2$ model produces
a total infrared luminosity, integrated from 8-1000 $\micron$
\citep{Kennicutt1998}, of $1.3 \times 10^{12}\ \rm L_{\sun}$ $\mu^{-1}$, where
$\mu$ is the magnification factor.  The dust temperature was determined to
be 36$\pm$1 K by fitting a modified blackbody to the peak of the dust bump with
$\beta$ fixed at 1.5, using Eq.  \ref{eq:modified-BB}.

\begin{equation}
S_{\nu} = N(\nu/\nu_{0})^{\beta}B_{\nu}(T)
\label{eq:modified-BB}
\end{equation}

\noindent B$_{\nu}$ is a Planck function (evaluated at single
temperature T), N is the amplitude, $\nu$ is the frequency, $\nu_{0}$ is the
frequency which is typically fixed at c/250 $\micron$, and $\beta$ is the
exponent that determines the shape of the modified blackbody. 

\begin{figure}
\begin{center}
\includegraphics[scale=0.45]{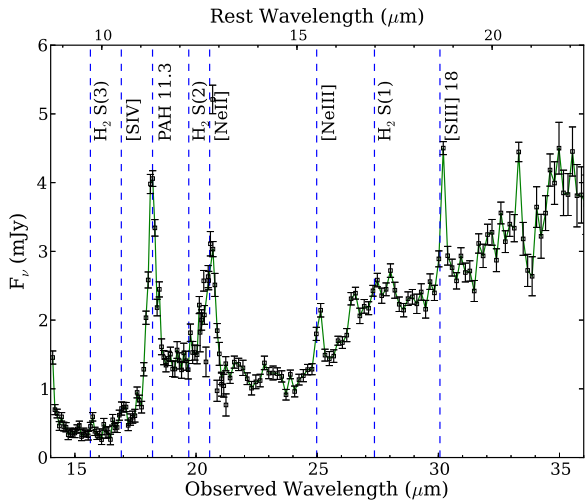}
\caption{\label{fig:irs_spectrum}The \spitzer/IRS spectrum of the lensed galaxy
AS1063a.  We observe the [Ne {\sc iii}] and [Ne {\sc ii}] lines which we
use to determine if our galaxy is undergoing AGN.}

\end{center}
\end{figure}

\begin{figure}
\begin{center}
\includegraphics[scale=0.61]{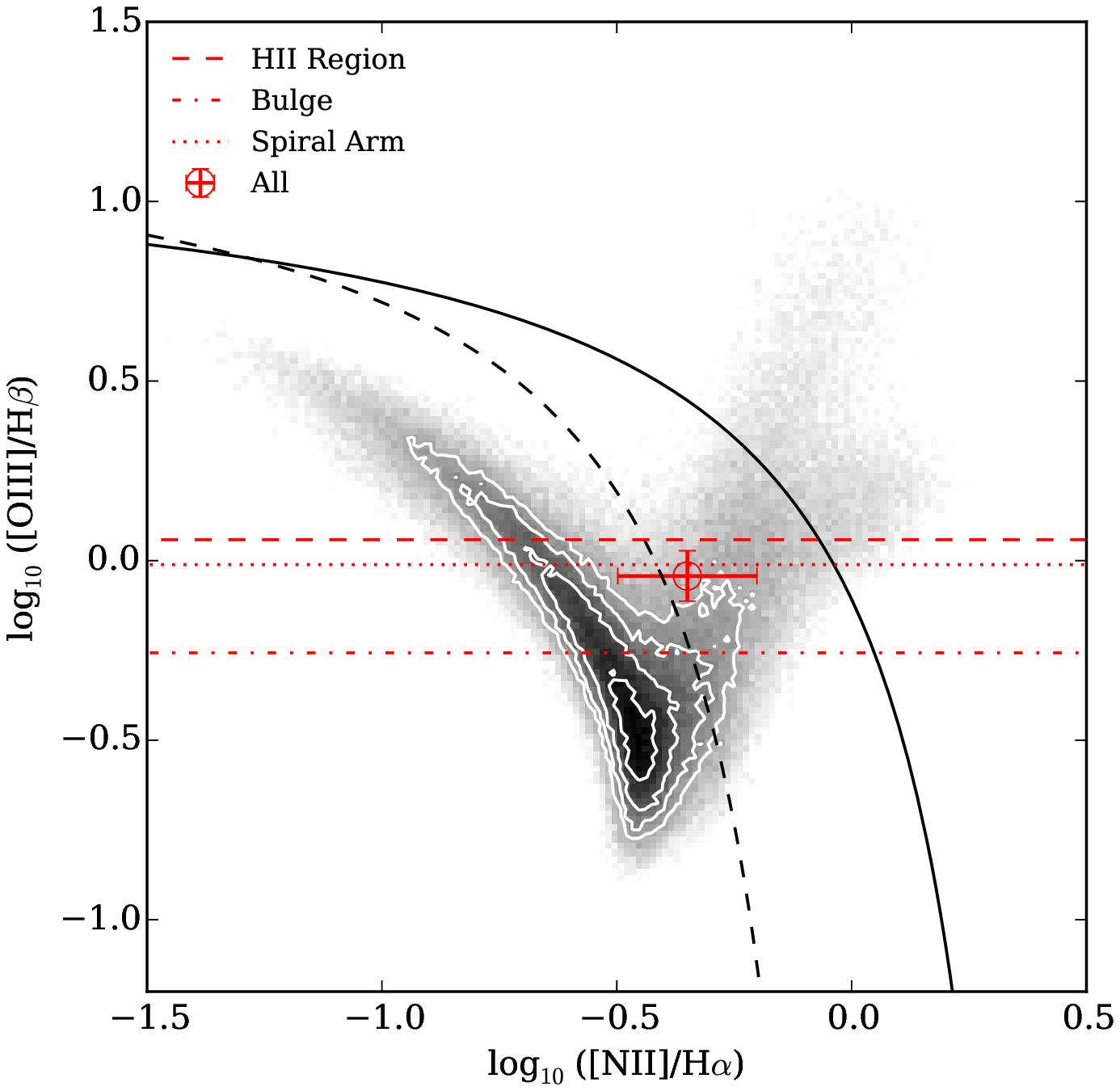}
\caption{\label{fig:BPT}The BPT diagram of the three regions within the
lensed galaxy AS1063a.  The grayscale points are a 2D histogram of SDSS
measured line ratios of redshifts between z = 0.005 - 0.25. The red lines
mark the [O {\sc iii}]/H$\beta$ ratios of three regions of the galaxy
constrained by the optical data.  Follow-up NIR observations in poor seeing
conditions roughly constrain the [N {\sc ii}]/H$\alpha$ ratio of the entire
galaxy.  The black lines mark the \citet{Kewley2001} and
\citet{Kauffmann2003} predictions for active galaxy type.  The error bars
include a 20\% systematic error in the flux calibration.}
\end{center}
\end{figure}

The \spitzer/IRS spectrum of this galaxy is shown in
Figure~\ref{fig:irs_spectrum}, and the measured line fluxes are listed in
Table~\ref{tab:irs_fluxes}.  The rest-frame mid-infrared spectrum looks like
that of a star-forming galaxy, with a strong PAH feature (11.3 $\mu$m).
The measured \neiii (15.5 $\mu$m)/\neii (12.8 $\mu$m) line ratio is less
than 1 (0.33$\pm$0.06), a value typical of a solar-metallicity starburst
galaxy \citep{Thornley2000, RigbyRieke2004}.  It is therefore clear
that the predominant source of the infrared-luminosity is star-formation
and not an AGN.  This is consistent with the fact that the measured optical
line ratios put this galaxy on the edge of the star forming
sequence/composite region in the
BPT diagram (Figure~\ref{fig:BPT})

\floattable
\begin{deluxetable}{lcr} 
\tabletypesize{\footnotesize}
\tablewidth{0pc}
\tablecaption{\label{tab:irs_fluxes}Mid-IR Line Fluxes of AS1063a}
\tablecolumns{3}
\tablehead{
\colhead{Line} & \colhead{Wavelength} & \colhead{Flux} \\
\colhead{} & \colhead{($\mu$m)} & \colhead{($10^{-17} {\rm erg/s/cm}^{2}$)}
}
\startdata
\ariii       &  9.0 &   7$\pm$15 \\
H$_{2}$ S(3) &  9.7 &  47$\pm$11 \\
\siv         & 10.5 &  62$\pm$13 \\
H$_{2}$ S(2) & 12.2 &  38$\pm$15 \\
\neii        & 12.8 & 238$\pm$16 \\
\neiii       & 15.5 &  78$\pm$13 \\
H$_{2}$ S(1) & 17.0 &  32$\pm$13 \\
\siii\ 18    & 18.7 & 167$\pm$ 8 \\
\enddata

\end{deluxetable}

\begin{figure*}
\begin{center}
\includegraphics[scale=0.9]{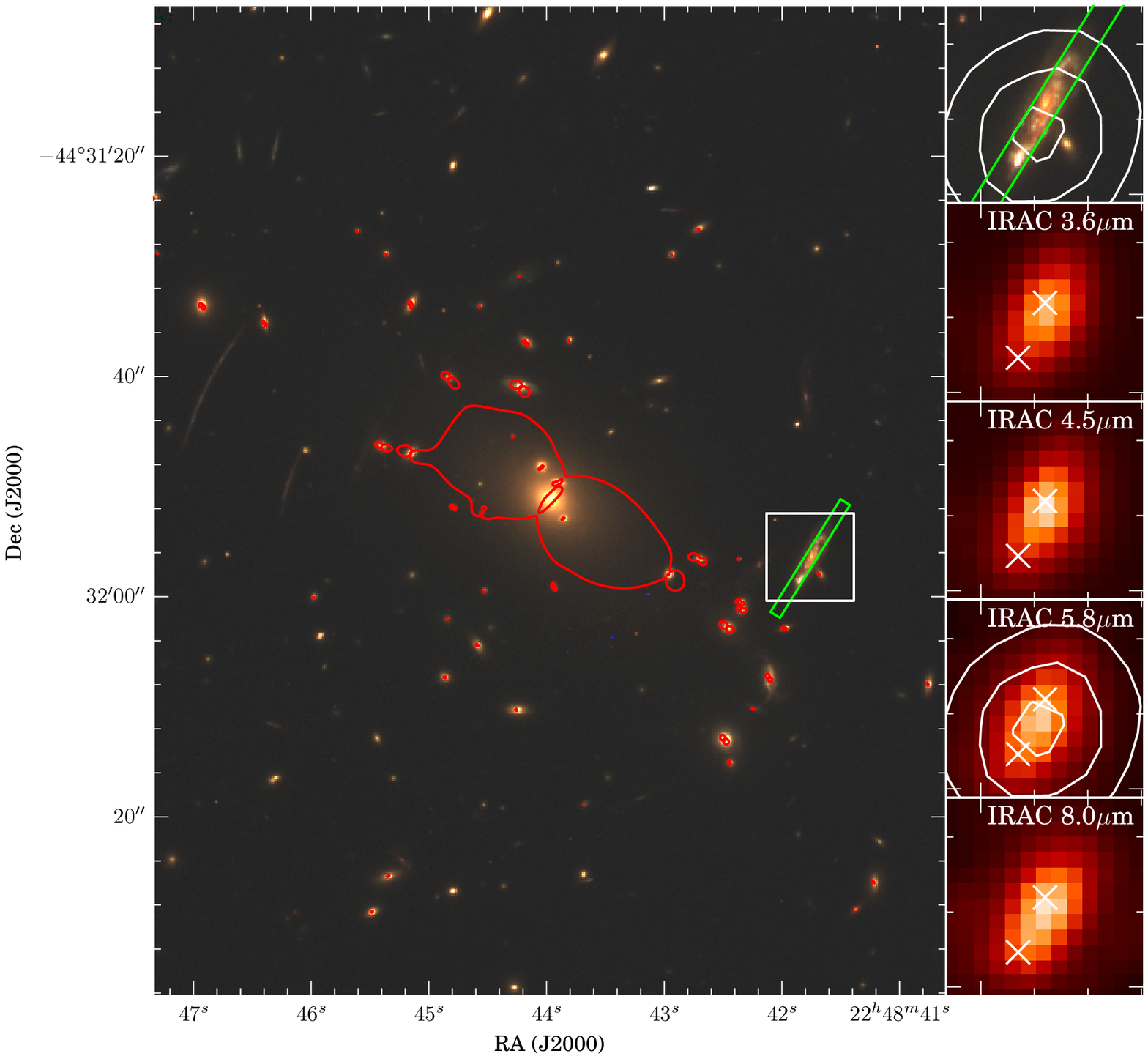}
\caption{\label{fig:hst_images}(left panel) Three color image of the cluster
core AS1063 using \hst\ ACS F606W, and \hst\ WFC3/IR F110W and F160W
filters. The main panel shows the central 1.0\arcmin$\times$1.2\arcmin\ area
of the cluster core. The red line marks the critical line at $z = 0.61$.  The
green box marks the position of the slit used for optical spectroscopy of
lensed galaxy AS1063a. The white box in the left panel shows the FOV of
the source AS1063a displayed in the right panels. (right panels) The top
right panel shows the zoomed \hst\ three color image, using the same filters
at the left panel.  The bottom four panels show each of the \spitzer/IRAC
3.6, 4.5, 5.8 and 8.0 $\micron$ images with the same FOV as the top right
panel. The white contour marks the position of the MIPS 24 $\micron$
contour.  White x's mark the positions of the bright star forming region and
bulge.} 
\end{center}
\end{figure*}

The \hst\ images show that this galaxy has a spiral-like morphology with
clearly defined bulge and disk components (Figure~\ref{fig:hst_images}).
In our optical spectrum obtained along the long axis of the galaxy
shows a velocity offset from one side to the other
(Figure~\ref{fig:ldss3_2d}), caused by the rotation of the disk component.
The proximity of the $z=0.61$ galaxy to the cluster center suggests that
its gravitational magnification is likely significant although its
normal-looking morphology of a spiral galaxy indicates that the
magnification effect is not large enough to destroy the intrinsic galaxy
morphology, essentially stretching the galaxy in the direction tangential
to the cluster center.

Figure~\ref{fig:ir_images} clearly shows that the brightest
infrared/submillimeter emission comes from this lensed galaxy at $z=0.61$.
It is, however, not clear which part of the galaxy is exactly responsible
for this strong infrared emission.  Figure~\ref{fig:hst_images} shows that the
peak of the brightest 24 $\mu$m emission is located 1.27\arcsec\ SSE from
the galaxy nucleus, falling in the middle of the disk.  No bright
optical/near-IR counterpart is seen at the peak of the 24 $\micron$ emission
although three smaller star forming clumps are seen nearby in the \hst\
images.  The 24 $\micron$ emission appears resolved in one axis, elongated
along the longer axis of the galaxy, suggesting the possibility that the 24
$\micron$ emission may originate from multiple components in the galaxy.

\subsubsection{Bright Optical Clump with Strong Line Emission}

What is most striking about the $z=0.61$ lensed galaxy, which corresponds
to the brightest MIPS 24 $\mu$m and PACS/SPIRE source, is its exceptionally
bright optical clump seen at one edge of the disk, 2.48\arcsec\ southeast
from the center of the galaxy.  The CLASH \hst\ data show that this clump
is the brightest feature in all of the \hst/ACS optical bands.  In the
\hst/WFC3 near-infrared bands, however, the central bulge becomes the
dominant feature.  

This bright optical clump appears to have a quite significant size
intrinsically.  By fitting an elliptical Gaussian to the \hst\ WFC3/IR F105W
image (which corresponds to rest-frame H$\alpha$) and subtracting in
quadrature a Gaussian point spread function (PSF) measured in the same
image, we have derived the spatial size of this clump as
0.547$\pm$0.025\arcsec $\times$ 0.099$\pm$0.034\arcsec.  This corresponds
to 3686$\pm$171 pc $\times$ 670$\pm$230 pc at $z=0.61$ if we do not take
into account the lensing effect. 
 
\begin{figure*}
\begin{center}
\includegraphics[scale=0.9]{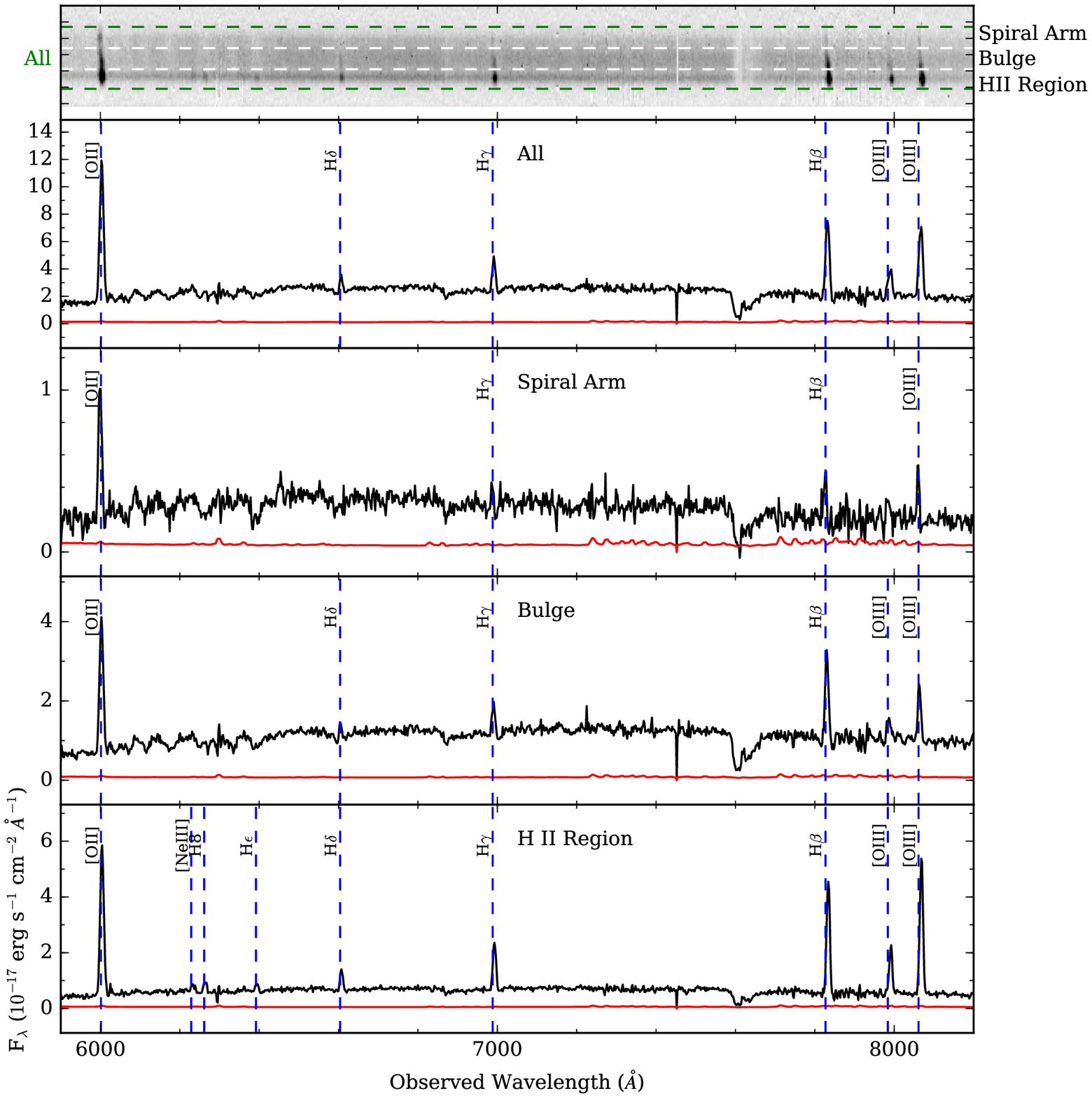}
\caption{\label{fig:ldss3_2d}(first panel) The 2D spectrum of the lensed
galaxy AS1063a taken with Magellan's LDSS-3.  (second panel) The 1D
spectrum of the entire galaxy.  (third panel) The 1D spectrum of the spiral
arm. There appears to be an older stellar population, the Balmer lines are
the weakest in this part of the galaxy.  (fourth panel) The 1D spectrum of
the bulge .  This region corresponds to the central region of the 2D
spectrum and the central region of the lensed galaxy. This region appears
to have a significant older stellar population.  (fifth panel) The 1D
spectrum of the H {\sc ii} region. This region corresponds to the lower
bright part of the 2D spectrum.  This is the bright star forming clump that
is dominated by strong Balmer and forbidden emission lines.} 

\end{center}
\end{figure*}

The 2D spectra of the $z=0.61$ galaxy shown in Figures
~\ref{fig:grism_spectra} and \ref{fig:ldss3_2d} show that the bright clump
emits strongly in line emission (e.g., Balmer lines H$\alpha$ - H8,
forbidden lines \oii\ $\lambda$3727, \oiii\ $\lambda\lambda$4959,5007,
\neiii, \sii\ and \siii).  Although we do detect the line emission
throughout the galaxy disk, it quickly becomes fainter away from the clump
while the Ca {\sc ii} H and K absorption features become more prominent,
reflecting an increasing light contribution from an older stellar
population in the bulge and spiral arm area on the other side of the galaxy
(Figure~\ref{fig:ldss3_2d}).  

\section{Discussion}

\subsection{Lens Model Reconstruction}
In order to determine the intrinsic properties of AS1063a it was necessary
to remove the effects of gravitational lensing. Modelling of the
gravitational lens, in the strong regime, was done with {\sc lenstool}
\citep{Kneib1996, Jullo2007, JulloKneib2009}.  {\sc lenstool} models the
cluster lenses by utilizing a non-parametric method, using multiple images
and redshifts of lensed background galaxies in order to constrain the
model.  In the newest CLASH data, we identified 5 multiply imaged systems,
one of which was confirmed with a spectroscopic redshift, that were used as
input constraints for the lens model \citep[Cl\'ement et al. in prep]{Richard2014}. With the lens
model it was possible to determine critical lines and spatial and flux
magnifications of the lensed galaxy. 

\begin{figure*}
\begin{center}
\includegraphics[scale=0.65]{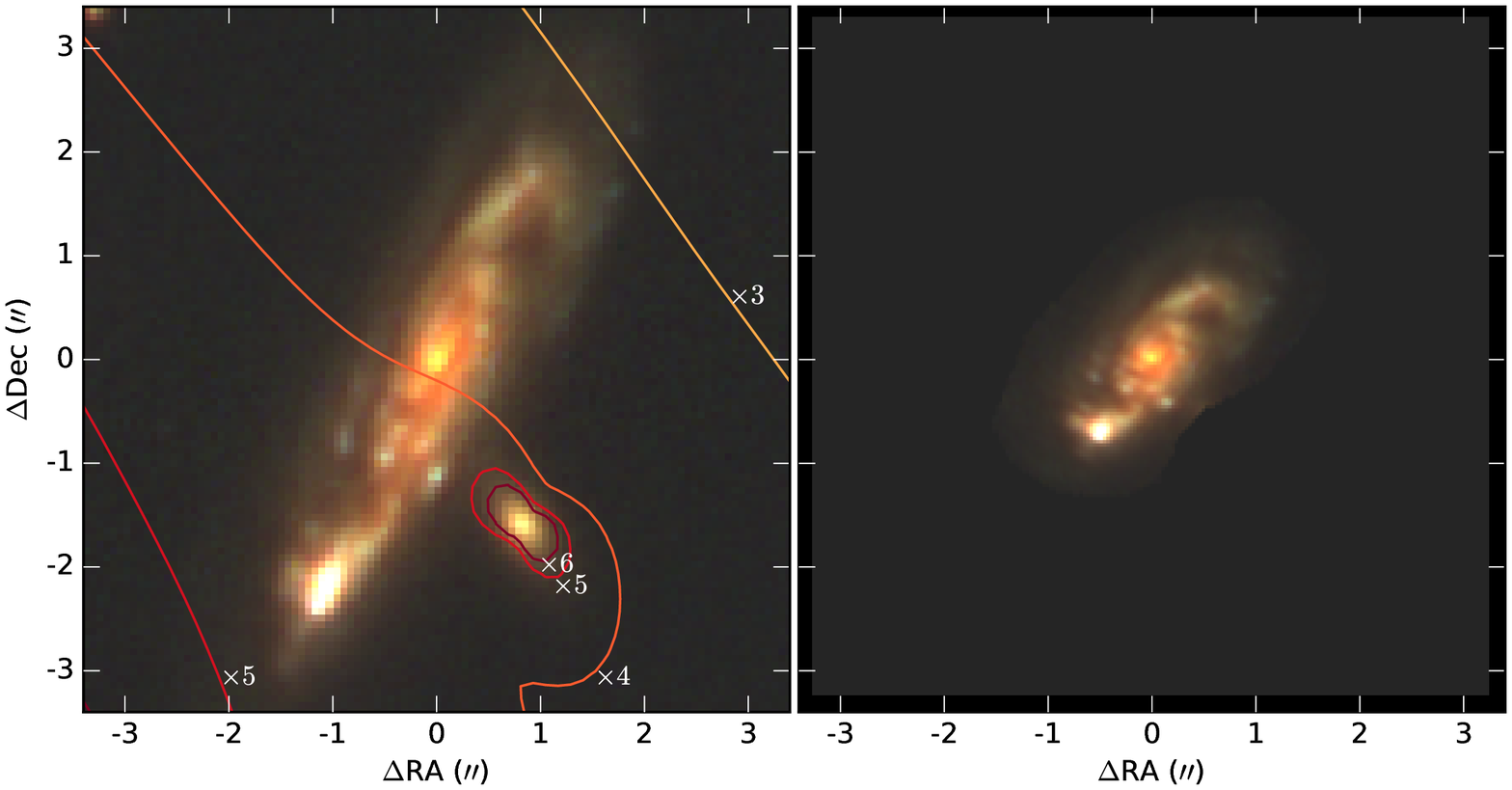}
\caption{\label{fig:source_plane}(left) \hst\ three color image (ACS F606W
and F814W, and WFC3/IR F110W filters) of the lensed galaxy AS1063a. The
contours represent the flux amplification map of the AS1063 lens. (right) The
source plane reconstruction performed by {\sc lenstool}.  Most of the
magnification is linear along the major axis of the galaxy. The morphology
of the lensed galaxy clearly is a spiral galaxy with two prominent spiral
arms.  The bright star forming clump is the dominant feature in all
ACS/WFC3 bands.}

\end{center}
\end{figure*}

From the lens model we are able to reconstruct AS1063a in the source plane
(Figure~\ref{fig:source_plane}, right panel).  Most of the magnification is
linear, with very little distortion.

The luminosity-weighted magnification is 4.0$\pm$0.1.  
 The error in the magnification is the statistical error and
does not include systematics such as choice in parameterization/modelling
or use of bad constraints/assumptions in the model.  AS1063a is
$\sim$5\arcsec\ from the critical line which suggests that is too far for
differential magnification to be a significant factor.

There is a slight gradient in the spatial distortion of the galaxy, from
the southern edge to the northern edge, 3.6$\times$ - 2.6$\times$ in linear
magnification.  In ACS this resolves objects larger than 90 - 130 pc, and
WFC3/IR 240 -- 340 pc.  The LDSS-3 spectrum, at 0.76\arcsec\ seeing, spatially
resolves 1.4 - 2.0 kpc.  The flux amplification from the southern edge to
the northern edge is 4.5$\times$ - 3.1$\times$, as shown in the right panel
in Figure~\ref{fig:source_plane}.  The bright star forming clump
corresponds to the region of the larger amplification at the southern edge
of the galaxy.
 
The bulge of AS1063a is prominent in the F775W filter and longward
wavelengths.  When comparing the observed image to the reconstructed image,
the galaxy only appears stretched in the observed image
(Figure~\ref{fig:source_plane}, left).  From the reconstructed image a few
noticeable features pop out; there are two spiral arms, a bulge and
multiple bright clumps (star forming knots) with one very prominent bright
clump (the giant H {\sc ii} region).  The magnification at the giant bright
clump is 4.44$\times$.

We discovered that there is also disagreement with the lens model
constructed by \citet{Gomez2012}.  A redshift was obtained for
``Lens B" and ``Lens C" as designated by \citet{Gomez2012}, which is
one of the main constraints for the non parametric lens model.
According to our lens model, based on the location of the bright arc,
AS1063a or ``Lens A", it is well outside the $z=0.61$
critical line and no additional images are expected for this arc. The
suggested counter image to ``Lens A", ``Counterpart A" is another
lensed arc.  With the additional resolution provided by the CLASH
data, it is clear that the colors do not match the colors of the bright
arc, and the bright clump seen in the ``Counterpart A" is not the
bright clump identified in the bright arc, but highly suggestive of a
cluster galaxy.

\subsection{Physical Properties of the $z=0.61$ Lensed IR-Luminous Galaxy}

\subsubsection{Star Formation Rates}

When corrected for a magnification factor of 4.0$\pm$0.1, the
intrinsic infrared luminosity of the $z=0.61$ galaxy becomes 
(3.1$\pm$0.1) $\times 10^{11}\ \rm L_{\sun}$ with a corresponding star
formation rate of 54$\pm$2 M$_{\sun}$ yr$^{-1}$.  This means that this
galaxy is intrinsically infrared-luminous and of the LIRG type (Luminous
Infrared Galaxy, L$_{\rm IR} \geq 10^{11}\ \rm L_{\sun}$.)

On the other hand, the star formation rate (SFR) derived from H$\alpha$, detected with MMIRS, is
significantly smaller.  The observed H$\alpha$ line luminosity gives a star
formation rate of 28$\pm$1 M$_{\sun}$ yr$^{-1}$.  Corrected for
magnification, the value decreases to 7.0$\pm$0.3 M$_{\sun}$
yr$^{-1}$.  When corrected for a visual extinction of A$_{\rm V} = $
1.5$\pm$0.2 mag (E(B-V) = 0.36$\pm$0.04) derived from the Balmer decrement
assuming case B recombination and using a Calzetti dust law, the value
increases to 33$\pm$1 M$_{\sun}$ yr$^{-1}$, but this is still smaller
than the infrared-derived value of 54$\pm$2 M$_{\sun}$ yr$^{-1}$.  This
implies that 40\% of the star formation is obscured.

The \hst\ grism spectrum of AS1063a is at a spatial resolution of
0.13\arcsec.  At this spatial resolution we can determine the contributions
for the spatially distinct components of the galaxy (i.e. \hii\ region,
bulge and spiral arm).  In order to utilize this information, we need to
assume an \nii/H$\alpha$ ratio.  From the MMIRS spectrum for the entire
galaxy, we measure a ratio of 0.45$\pm$0.03, which we assume for each of
the regions in the galaxy.  Unfortunately, the seeing and S/N of the MMIRS
spectrum is not sufficient enough to determine the ratio for the spatially
distinct components of the galaxy.  We also assume that the ratio between
\nii$\lambda$6583/\nii$\lambda$6548 $\sim$ 3.  With this we estimate the
H$\alpha$ flux for the entire galaxy as $340.9\pm26.3\times10^{-17}$
ergs s$^{-1}$ cm$^{-2}$; 171.9$\pm$13.3 ergs s$^{-1}$ cm$^{-2}$ for the \hii\ region,
128.2$\pm$9.9 ergs s$^{-1}$ cm$^{-2}$ for the bulge and 31.6$\pm$2.5
ergs s$^{-1}$ cm$^{-2}$ for the spiral arm.  Using the E(B-V) computed for the
entire galaxy and correcting for the magnification of each of the regions
(i.e. 4.4$\times$, 3.9$\times$, and 3.3$\times$), we find that the SFR
for each of the regions are the following: 14$\pm$1 M$_{\sun}$
yr$^{-1}$ for the \hii\ region, 12$\pm$1 M$_{\sun}$ yr$^{-1}$ for the
bulge, and 4$\pm$1 M$_{\sun}$ yr$^{-1}$ for the spiral arm.

In \S\ref{sec:stellar_mass}, E(B-V) is also computed from the SED fitting
of the photometry for AS1063a and its \hii\ region, which are 0.31 and 0.18
respectively.  If we assume these E(B-V) values and magnification
corrections, it would affect the \hst\ grism based SFRs in the following
way: 27$\pm$2 M$_{\sun}$ yr$^{-1}$ for AS1063a and 8$\pm$1
M$_{\sun}$ yr$^{-1}$ for the \hii\ region.  The overall E(B-V) computed for
the galaxy from the Balmer decrement agrees with the one computed from the
photometry.  However, for the \hii\ region, it might be expected that there
could be some geometry or sight-line effect that could result in a
different E(B-V) value. In particular, lower in this case, which is evident
by the \hii\ region being the brightest feature in the galaxy in the
UV-optical bands as well as detecting more Balmer transitions (i.e.
H$\alpha$-H8). For the \hii\ region we adopt the SFR = 8$\pm$1 M$_{\sun}$
yr$^{-1}$ that is derived using the E(B-V) value from the photometry, which
seems more consistent with the evidence.

Using H$\beta$ and H$\gamma$ and H$\delta$ lines seen in the 2D optical
spectrum (Figure~\ref{fig:ldss3_2d}), we can derive visual extinctions and
extinction-corrected star formation rates for spatially distinct components
in the galaxy. However, the E(B-V) values disagree with the one derived
from H$\alpha$ and H$\beta$, increasing with higher order Balmer
transitions.  Upon further inspection it appears that self absorption from
the continuum could be affecting the flux measured for the higher order
Balmer lines.

\subsubsection{Precise Location of the Infrared Source}

\label{sec:ir_location}

As we mentioned in \S\ref{sec:ir_luminous_galaxy} the MIPS 24 $\micron$
emission is offset from the bulge of the $z=0.61$ galaxy by 1.27\arcsec and
elongated, suggesting it is resolved in one axis.  One concern is whether
the offset is real or if it is a result of astrometric error due to the
large MIPS 24 $\micron$ PSF. The native pixel scale of the MIPS 24
$\micron$ is 2.45\arcsec\ pixel$^{-1}$. However, the image we use is sampled at
1.245\arcsec\ pixel$^{-1}$.   In order to compare the bulge position to the MIPS
position we need a set of unresolved sources that can be used for
astrometry. For the $z=0.61$ lensed galaxy, the bulge can be identified in
the F814W filter, which is also convenient due to the larger FOV of ACS.
We used two methods to identify stars in the \hst\ F814W images; 1)
plotting the difference in the magnitude of two different sized apertures
and 2) plotting magnitude versus half-light radius at a fixed aperture.
Stars will resemble the PSF of the instrument for a wide range of radii,
whereas if just using a FWHM measure, some galaxies may have similar FWHM
as a star and lead to a false positive detection. As an additional check,
we visually inspected stars to ensure sources were not misidentified and
were not blended with other sources. For sources we believe to be stars, we
checked to see whether they are detected in IRAC and also to ensure they
are not blended.  From our clean list of point sources, we compute the
separation angle between the sources detected in ACS and IRAC. We then
measure the standard deviation of all the separations, computing the
1$\sigma$ error in the astrometric position of 0.12\arcsec.

At 24 $\micron$ many of the sources are unresolved, very few of which are
actually stars.  In order to determine the astrometric error between IRAC
and MIPS we rely on compact sources that are unresolved in IRAC.  We use a
similar strategy to the one described above to select point sources in
IRAC. We compute the standard deviation of the separation angle between the
positions in IRAC to the positions in MIPS and find an error of
0.23\arcsec.  The combined 1$\sigma$ error in the astrometric positions in
ACS, IRAC and MIPS is 0.26\arcsec.  At 4.9$\sigma$ we can confidently say
that the offset of the MIPS peak emission is real. 
 
In order to explain the elongation of the source in MIPS 24 $\micron$, we
simulate the source by convolving a Gaussian with the MIPS PSF.  We then
assume the position is at the center of the peak MIPS emission and vary the
amplitude while minimizing $\chi^{2}$.  We find that a single source at the
peak of the 24 $\micron$ emission is not sufficient to account for all the
flux in the source, which has significant residual wings.  To account for
all the flux it appears that more than one Gaussian is needed.  We next
assume that the far-IR emission could be coming from two locations, offset
from the peak emission. Two reasonable locations are the bulge of the
galaxy and the star forming region.  We find that fitting two Gaussians at
those locations works well. The residuals are much smaller with the
majority of the flux being accounted for.  The solution that works the best
is two Gaussians with roughly the same amplitude.  However, these solutions
are highly degenerate and additional Gaussians could potentially be added
for better fits.  We can confidently say that it is unlikely that a single
source is responsible for all of the far-IR emission at the position of the 24
$\micron$ emission peak. There must be more than one source
contributing to the far-IR flux; higher spatial resolution is needed to
resolve the situation.

Looking at the IRAC bands, there is a slight bump at 5.8 $\micron$.  Based
on the redshift of this galaxy, it is expected that the 3.3 $\micron$ PAH
feature would be completely in the band. However, the 3.3 $\micron$ PAH
feature is much weaker than the other PAH features found in star forming
galaxies and the bump we see in the SED may not be caused by this feature.
It is expected that the majority of light contributing to the flux in the
IRAC bands is from stars. If there is flux coming from the 3.3 $\micron$
PAH feature, we would need to subtract off the stellar light.  To account
for the stellar light, we scale the flux in IRAC 3.6 to the IRAC 5.8
$\micron$ image and subtract them from each other.  We also do the same
with the IRAC 4.5 $\micron$ image. In each case there appears to be no
evidence for residual flux coming from the 3.3 $\micron$ PAH feature. If
the bump seen in the SED is real, it is possible that the PAH emission
could be throughout the galaxy at a low enough level that it might not be
seen in the residuals.  It appears unlikely that the flux would be
originating from an individual region in the galaxy, otherwise it should
have been seen in the residuals.

\subsubsection{Stellar Mass}
\label{sec:stellar_mass}

\floattable
\begin{deluxetable}{lccc}
\tabletypesize{\footnotesize}
\tablewidth{0pc}
\tablecaption{\label{tab:stellar_mass}Stellar Mass of AS1063a}
\tablecolumns{4}
\tablehead{
\colhead{Metallicity} & \colhead{A$_{v}$} & \colhead{Age} & \colhead{Stellar mass} \\
\colhead{[$Z_{\sun}$]} & \colhead{} & \colhead{[Gyr]} & \colhead{[$M_{\sun}$]} \\
}
\startdata
\multicolumn{4}{c}{Entire galaxy} \\
\hline
1.000 & 1.25$^{+0.21}_{-0.48}$ & 0.65$^{+0.48}_{-0.27}$ & 2.6$^{+1.5}_{-1.0}\times10^{10}$ \\
0.005 & 1.61$^{+0.18}_{-0.19}$ & 0.64$^{+0.53}_{-0.21}$ & 3.0$^{+1.7}_{-1.1}\times10^{10}$ \\
\hline
\multicolumn{4}{c}{H {\sc ii} region} \\
\hline
1.000 & 0.74$^{+0.20}_{-0.24}$ & 0.38$^{+0.31}_{-0.17}$ & 10.5$^{+6.1}_{-3.9}\times10^{8}$ \\
0.200 & 1.15$^{+0.30}_{-0.41}$ & 0.10$^{+0.16}_{-0.07}$ &  7.1$^{+4.1}_{-2.6}\times10^{8}$ \\
\enddata

\end{deluxetable}

We fit the \hst\ 16-band photometry to \citet{BruzualCharlot2003} models
following \citet{Perez-Gonzalez2008,Perez-Gonzalez2013} in order to
determine stellar mass, dust attenuation, and stellar age and population.
We assume a Chabrier IMF, Calzetti extinction and exponentially decreasing
star formation history.  The models were run fixing the metallicity to
solar metallicity and also allowing it to be a free parameter. The SED
fitting was conducted on the photometry of AS1063a and just the H {\sc ii}
region.  The images used for the photometry of the H {\sc ii} region were
PSF matched to ensure that we are comparing the same physical region in
each band.  We used {\sc TinyTim} \citep{Krist2011} to generate the PSF for each of the 16
bands, convolved with the charge diffusion PSF and rebinned to the CLASH
pixel scale of 65 mas.  The {\sc iraf} task \verb'psfmatch' was used to
convolve all of the \hst\ images to the WFC3/F160W PSF. 

\begin{figure*}
\begin{center}
\includegraphics[scale=0.45,angle=-90]{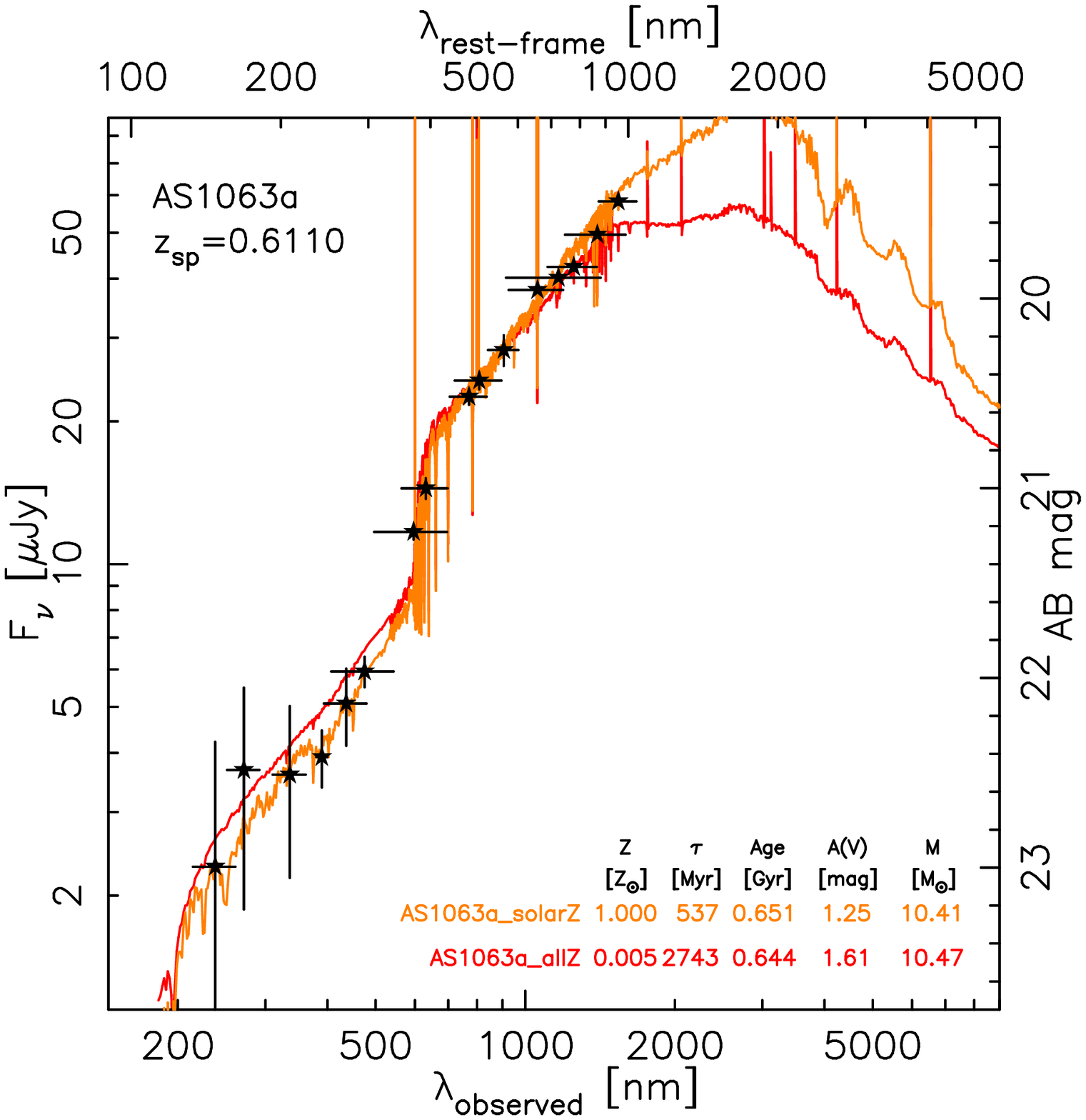}
\hspace{1cm}
\includegraphics[scale=0.45,angle=-90]{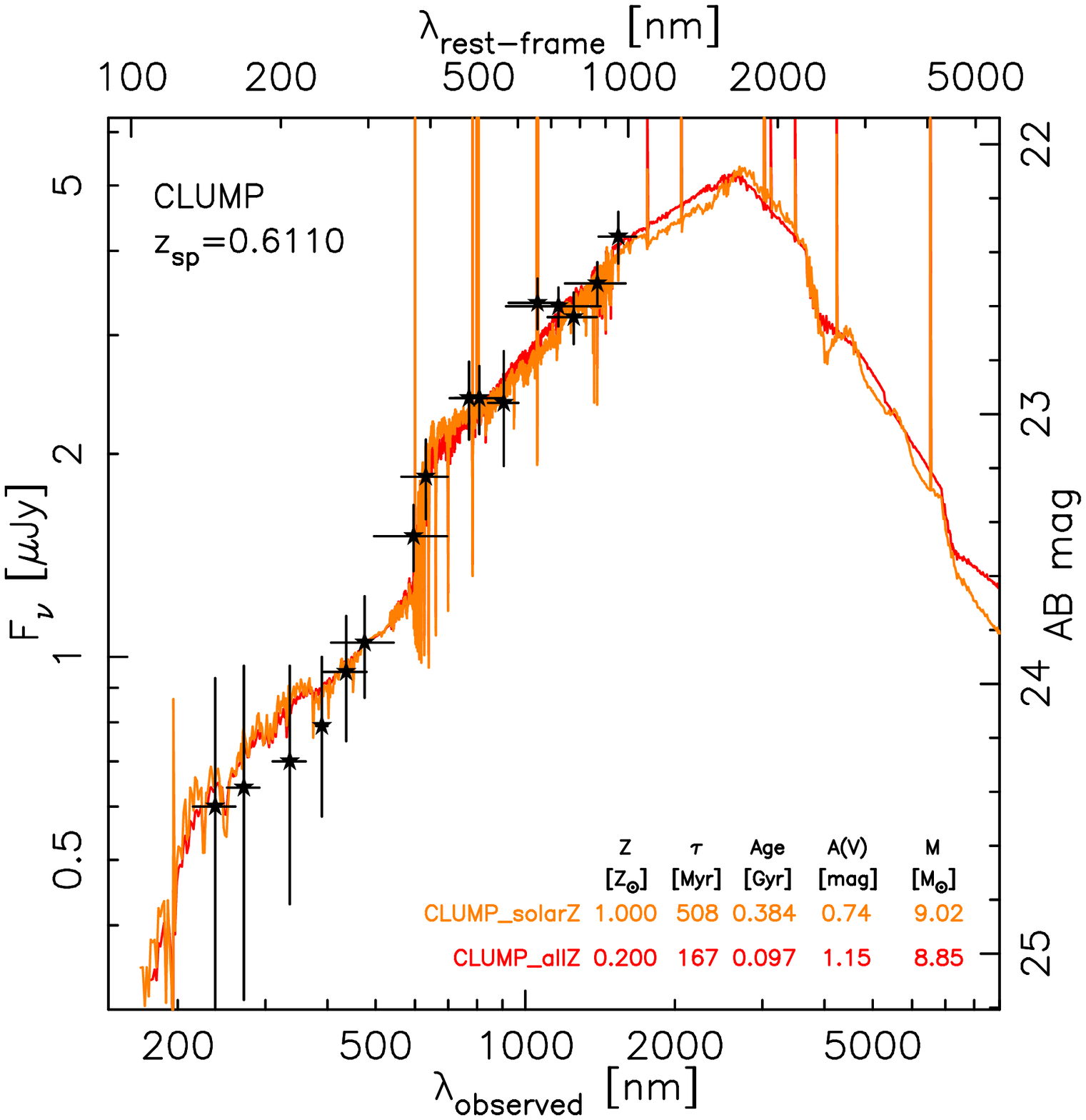}
\end{center}
\caption{\label{fig:hst_phot}Left: SED fits to the intrinsic flux of the
entire galaxy, AS1063a.  Right: SED fits to the intrinsic flux of \hii\
region in AS1063a.  The templates used assume a Chabrier IMF, Calzetti
extinction and exponentially decreasing star formation history. The orange
line shows solar metallicity template and the red line is the best fitting
metallicity. From the photometry fitting, for both AS1063a and the clump,
neither model is preferred.  The metallicity determined from the nebular
emission lines from the R23 and N2 diagnostics help constrain which model
is the appropriate one, that being the solar metallicity model. } 
\end{figure*}

The results of the SED fitting can be found in Table \ref{tab:stellar_mass}
and Figure \ref{fig:hst_phot}.  The stellar masses derived from the SED
fitting are good to within 0.2 dex, including uncertainties in the IMF and
other systematics.  AS1063a's stellar mass appears to be unaffected by
metallicity as its mass varies by 1.1$\times$ for the metallicities
considered.  In \S\ref{sec:metallicity} we describe the metallicity derived
from the nebular emission lines in which we find the value is about solar
metallicity.  For AS1063a we adopt the stellar mass (2.6$\times10^{10}\ \rm M_{\sun}$) at solar metallicity.  The \hii\ region on the other hand is
seems to vary by a factor of 1.5$\times$ depending on the metallicity.  We
also adopt the solar metallicity value (1.1$\times10^{9}\ \rm M_{\sun}$), in
which it appears to be close to solar metallicity, as we go over in
\S\ref{sec:metallicity}.

When comparing the H {\sc ii} region in AS1063a to ones at higher redshift
we find that is about average stellar mass, the high redshift H {\sc ii}
regions span $10^{7}$ -- $10^{10}\ \rm M_{\sun}$.  Looking at the samples
individually; the \hii\ regions in RCSGA0327 \citep{Wuyts2014}, a galaxy a
$z = 1.7$, fall on the lower mass side from $10^{7}$ -- $10^{8}\ \rm M_{\sun}$
whereas the \citet{Wisnioski2012} and \citep{ForsterSchreiber2011b} samples
are on the higher mass range $10^{9}$ -- $10^{10}\ \rm M_{\sun}$ and $10^{8}$
-- $10^{10}\ \rm M_{\sun}$.  This can be mostly explained by
\citet{Wisnioski2012} and \citet{ForsterSchreiber2011b} samples being
unlensed, so only the largest most massive H {\sc ii} regions are probed,
whereas the \citet{Wuyts2014} sample is a single lensed galaxy probing
small H {\sc ii} regions at high spatial resolution.

Using the stellar mass at solar metallicity for AS1063a and the SFR derived
from the far-IR, we compute the specific star formation
rate (sSFR), which is 2.11 Gyr$^{-1}$.  We use the analytic fitting
function from \citet{Whitaker2012} to determine the sSFR of the
star-forming main sequence, which is 0.35 Gyr$^{-1}$ for a galaxy with the
same stellar mass and redshift as AS1063a.  Galaxies with sSFRs 4$\times$
the star-forming main sequence are considered starbursts
\citep{Rodighiero2011, Noeske2007}.  The starburst region for a galaxy
similar to AS1063a is 1.38 Gyr$^{-1}$, which means that AS1063a
is a starburst.




\subsubsection{Kinematics}

\label{sec:kinematics}

\begin{figure*}
\begin{center}
\includegraphics[scale=0.7]{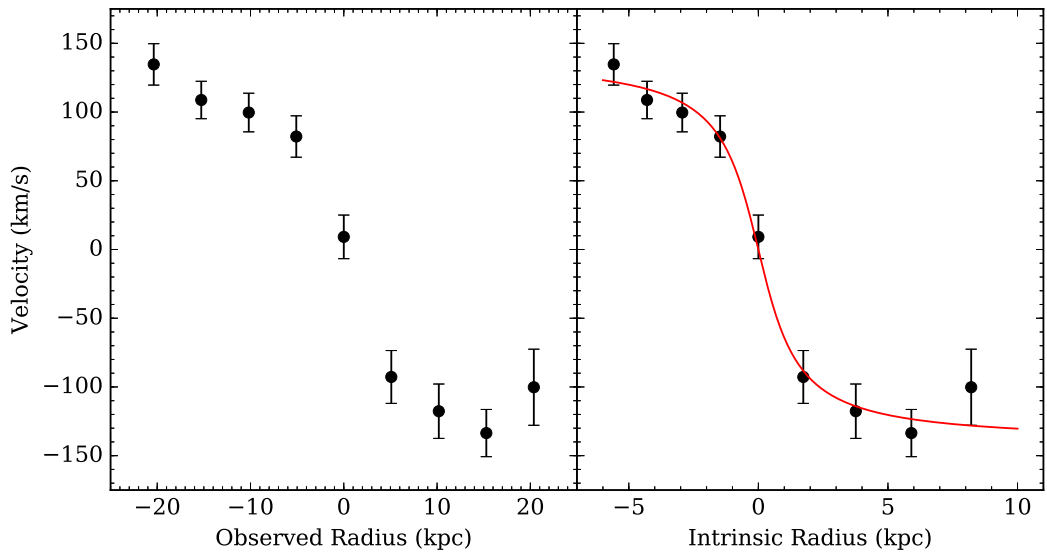}
\caption{\label{fig:kinematics} (left) The kinematics of the observed \oii\ emission line measured
from the 2D spectrum.  We measured the centroid of the blended
\oii$\lambda\lambda$3726,3729 doublet, which corresponds to \oii\
$\lambda$3727. \oii\ is the brightest feature spatially with the highest
signal-to-noise ratio.  In order to reduce the noise in the measurement, we
binned the data into bins of 4 pixels each (0.76\arcsec), which is
comparable to the seeing.  (right) The intrinsic kinematics of the lensed
galaxy AS1063a.  The red line is the best fit velocity profile.  The galaxy
appears to have a slight inclination which is seen by a velocity gradient,
suggesting that the galaxy is rotating as one system.  The rotation curve
appears to be flattening out outside 2 kpc.} 

\end{center}
\end{figure*}

Spectroscopically, this bright clump appears like an \hii\ region embedded
in a rotating disk, with a rotational velocity V$_{\rm max}$ of 134$\pm$17
km s$^{-1}$ at $R = 5.9$ kpc, which is also found by \citet{Gomez2012} and
\citet{Karman2015}.  The clump itself shows strong Balmer emission
(H$\alpha$ - H8), forbidden line emission ([O {\sc ii}]$\lambda$3727, [O
{\sc iii}]$\lambda\lambda$4959,5007, [Ne {\sc iii}]) emission, and its
systemic velocity falls on the rotation curve.  AS1063a has a smooth
rotation curve that flattens out outside of 2 kpc
(Figure~\ref{fig:kinematics}).

A system in the local Universe with a similar appearance to AS1063a is NGC
5194 (M51), which is a spiral galaxy interacting with another galaxy (NGC
5195).  \citet{Sofue1999} looked at the rotation curves of local galaxies
and found that NGC 5194 had a peculiar rotation curve (non-Keplerian).
NGC3034 (M82), a starburst galaxy interacting with M81, also shows a
peculiar rotation curve, in which it has a steep Keplerian decline.  In the
local Universe it is expected that an isolated spiral galaxy should have a
flat rotation curve at large radii (5-30 kpc).  Significant bumps or
deviations from Keplerian rotation in the rotation curve could indicate an
interaction with another galaxy or underlying substructure (subhalos).
Kinematically it is difficult to completely rule out AS1063a as a merger,
however it appears unlikely. 

From the rotation curve we can also measure a dynamical mass.  In order to
compute an accurate dynamical mass we need to determine the inclination of
the galaxy.  Using the source plane reconstruction of the galaxy, we measure
the axis ratio (b/a) using {\sc galfit} \citep{Peng2002,Peng2010}.   We find an axis ratio of b/a =
0.502$\pm$0.007.  Assuming the galaxy is an oblate spheroid
\citep{Holmberg1958}, we use equation \ref{eq:inclination} to compute the
inclination,

\begin{equation}
{\rm cos}^{2} i = \frac{(b/a)^{2} - q^{2}}{1-q^{2}}
\label{eq:inclination}
\end{equation}

\noindent where $i$ is the inclination angle, $b/a$ is the axis ratio
($a$ is the semi-major axis, $b$ is the semi-minor axis of the galaxy), and
$q$ is the axis ratio for an edge-on galaxy.  Typically $q$ is 0.13 in spiral
galaxies.  In order to understand the uncertainty in the inclination
angle we performed an MCMC using 1000 realizations of the AS1063 lens model
({\sc lenstool} routine \verb'bayesCleanlens'), with each realization
reconstructing AS1063a and the \hst\ PSF in the source plane.  We then use
{\sc galfit} to fit each of the realizations of the reconstructed source
plane image, using the reconstructed PSF associated with a reconstructed
image.  The standard deviation of the PA of the galaxy was 1.5$^{\circ}$.
Even with this deviation in the PA, we find that the uncertainty in the
inclination is 0.5$^{\circ}$. When we vary the parameter $q$, the disc
thickness from 0.11--0.2 \citep{Courteau1997}, which is typical for spiral
galaxies, we find the that it varies by 1.5$^{\circ}$. Given the systematic
uncertainties for determining inclinations of galaxies (e.g.
\citet{BarnesSellwood2003}), we adopt $\pm$5$^{\circ}$ uncertainties on the
inclination. We compute an inclination $i = 61\pm5^{\circ}$.  The
dynamical mass is defined by Eq. \ref{eq:mdyn},

\begin{equation}
M_{\rm dyn} = \frac{R V^{2}}{G}
\label{eq:mdyn}
\end{equation}

\noindent where $V = V_{r}\ sin(i)$ ($V_r$ is the radial velocity, $i$
is the inclination) and $R$ is the radius.  \citet{ForsterSchreiber2009}
uses the velocity at 10 kpc which is typically the surface brightness limit
of their sample, to determine the dark matter contribution of their
galaxies.  Looking at the rotation curves of nearby galaxies
\citep{Sofue1999}, many of the rotation curves flatten out well beyond 10
kpc, which means that the baryons contribute more to the inner 10 kpc than
dark matter.   For AS1063a, the rotation curve flattens out after 4-5 kpc
and it is not unreasonable to extrapolate the velocity at 10 kpc.
\citet{Courteau1997} demonstrates that the arctangent function is a good fit
to galaxy rotation curves in the local Universe in order to determine the
circular velocity at a given radius, shown in Eq. \ref{eq:arctan}, 

\begin{equation}
V(R) = V_{0} + \frac{2}{\pi} V_{c}\ {\rm arctan} \frac{R}{R_{t}}
\label{eq:arctan}
\end{equation}

\noindent where $V_{0}$ is the velocity center of rotation (systemic
velocity), $V_{c}$ is the asymptotic velocity, and $R_{t}$ is the  scale
radius where the rotation curve begins to flatten out. We fit the AS1063a
velocity curve with the arctangent function.  Figure \ref{fig:kinematics}
(right panel) shows the fit to the intrinsic radius.  The large error bars
and scatter seen in the points that are furthest from the center are due to
the low S/N of the fit to the \oii\ line.  If you allow the \oii\ line to be
fit at larger radius, the error bar increases significantly, which once
again reflects the S/N decreasing.

With the inclination and the circular velocity at 10 kpc, we compute a
dynamical mass of (5$\pm$1)$\times10^{10}$ M$_{\sun}$.  This dynamical mass is
comparable to the average SINS \citep{ForsterSchreiber2009} galaxy, which
spans $\sim$1--20 $\times10^{10}$  M$_{\sun}$ for redshifts $z = 1.3 - 2.6$.

%
%

\subsubsection{Metallicity}

\label{sec:metallicity}

\floattable
\begin{deluxetable}{lcc}
\tabletypesize{\footnotesize}
\tablewidth{0pc}
\tablecaption{\label{tab:metallicity}Metallicity of AS1063a}
\tablecolumns{3}
\tablehead{
\colhead{}    & \colhead{$12+log([{\rm O/H}])$} & \colhead{$12+log([{\rm
O/H}])$} \\
\colhead{}    & \colhead{R23}             & \colhead{N2} \\
}
\startdata
Entire galaxy & 8.95$\pm$0.02  & 8.80$\pm$0.03 \\
\hline                           
\hii\ region  & 8.96$\pm$0.02  & \nodata \\
Bulge         & 8.99$\pm$0.02  & \nodata \\
Spiral arm    & 8.74$\pm$0.10  & \nodata \\
\enddata

\end{deluxetable}

As mentioned in \S\ref{sec:ir_luminous_galaxy} we detect key emission lines
with LDSS-3 and MMIRS for determining metallicity. We use both R23 ([O {\sc
ii}]$\lambda$3727, H$\beta$, and [O {\sc iii}]$\lambda$5007) and [O {\sc
iii}]/[O {\sc ii}], and N2 (H$\alpha$ and [N {\sc ii}]$\lambda$6585)
diagnostics to determine the metallicity of the lensed galaxy at $z=0.61$
and its individual regions.  We were unable to detect the [O {\sc
iii}]$\lambda$4363 line, a low metallicity indicator, which implies that we
should choose the upper branch for R23.

For the N2 measurement we follow the prescription of
\citet{PettiniPagel2004} using their cubic relation.  For the R23
measurement, we follow the prescription of \citet{KewleyEllison2008},
specifically following the KK04 method.  The results for the metallicities
can be found in Table \ref{tab:metallicity}.  All the values are roughly
consistent with solar metallicity.  There is some evidence of a metallicity
gradient from the spiral arm to the bulge.  The metallicity is constant
from the bulge to the H {\sc ii} region.

\begin{figure}
\begin{center}
\includegraphics[scale=0.45]{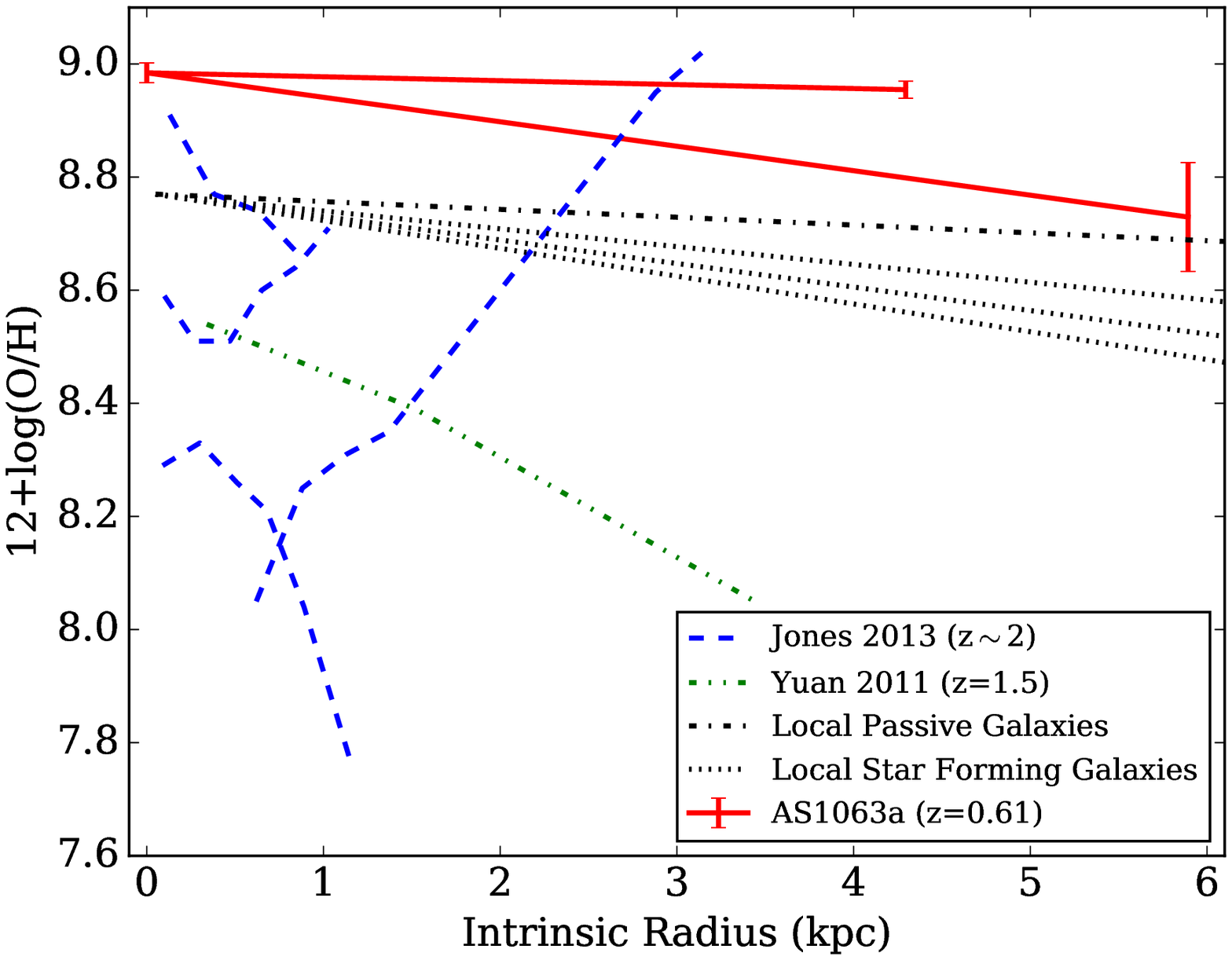}
\caption{\label{fig:metallicity}
Metallicity as a function of radius for galaxies in the local Universe
\citep{HenryWorthey1999, Rupke2010} and at high redshift gravitationally
lensed sample \citep{Yuan2011, Jones2013}. The higher redshift galaxies show
a steeper metallicity gradient than seen in the local Universe. AS1063a
resembles more local star forming galaxy.  We use the R23 to derive the
metallcity in AS1063a. \citet{Jones2013} uses a variety of methods, we
compare to their O3HB. \citep{Yuan2011} uses N2.} 
\end{center}
\end{figure}

Metallicity gradients are seen in the Milky Way, where the gas-phase metallicity is lower at larger radii 
 than at the bulge.  These metallicity gradients are also seen in local
galaxies, where the slope of the gradient is shallow.   The gradient in
metallicity suggests that the metallicity throughout the disk of a galaxy
is evolving over time.  At higher redshift, in $z\sim2$ galaxies
\citep{Jones2010b, Yuan2011, Jones2013}, there is evidence for steeper metallicity
gradients.  It is suggested that there is inside-out growth in these
galaxies that may be responsible for creating the steep metallicity
gradients.  In AS1063a, the metallicity gradient is not as steep as seen in
$z\sim2$ galaxies, and appears more like galaxies in the local Universe,
shown in Figure \ref{fig:metallicity}.

There are known aperture effects when measuring metallicity, where low
metallicity regions may be hidden by higher metallicity regions.  While it
appears that the metallicity is constant between the H {\sc ii} region and
the bulge of AS1063a, we may not be sensitive to the lower metallicity
regions.  This is primarily due to the spatial resolution of the spectra,
determined by the seeing, and S/N of the fainter emission lines, such as
the [O {\sc iii}]$\lambda$4959.

The SINGS sample of H {\sc ii} regions \citep{Moustakas2010} in the local
Universe spans metallicities 7.7 -- 9.3 (KK04).  The majority of H {\sc ii}
regions (88\%) have metallicities between 8.6 -- 9.2.   The metallicity of
the H {\sc ii} region in AS1063a does not seem unusual when compared to the
SINGS galaxies in the local Universe as 48\% of their H {\sc ii} regions
have metallicities between 8.9 -- 9.1.  Higher redshift  H {\sc ii} regions
seem to have lower metallicities when compared to the H {\sc ii} region in
AS1063a. \citet{Wisnioski2012} metallicities span 8.4 -- 8.8,
\citet{Jones2010b,Jones2013} span 8.2 -- 9.0. and \citet{Wuyts2014}
metallicities spans 8.0 -- 8.3 for several H {\sc ii} regions in RCSGA0327.

\subsubsection{Gas Depletion Timescale}

In order to estimate the molecular gas mass without a CO measurement there
are two methods; (1) using the dust mass from the galaxy's far-IR emission and
(2) using the dynamical mass from the galaxy's kinematics.

For the first method in determining gas mass, we need to measure the dust mass.
To determine the dust mass from the far-IR emission, we use Eq.
\ref{eq:dust_mass} \citep{Greve2012},

\begin{equation}
M_{\rm dust} = \frac{D_{L}^{2} S_{\nu_{o}}}{(1+z)\kappa_{\nu_{r}}}
[B_{\nu_{r}}(T_{\rm dust}) - B_{\nu_{r}}(T_{\rm CMB}(z))]^{-1}
\label{eq:dust_mass}
\end{equation}

\noindent where $S_{\nu_{o}}$ is the observed flux, B$_{\nu_{r}}$ is
the rest-frame Planck function, $\kappa_{\nu_{r}}$ is the rest-frame
absorption coefficient. The Planck function is defined at the dust
temperature T$_{\rm dust}$ and the  T$_{\rm CMB}$ at redshift z.  The
absorption coefficient is defined as $\kappa_{\nu_{r}} = 0.45(\nu_{r}/250\
\rm GHz)^{\beta}$ \citep{Hildebrand1983,KruegelSiebenmorgen1994}, where
$\nu_{r}$ is the rest-frame frequency and $\beta=2.0$.  According to
\citet{Papadopoulos2000} the temperature of the CMB affects the dust mass by
$\sim$2\%.  \citet{daCunha2013} also found that at redshifts $z<4$ the CMB
temperature negligibly affects the dust mass and because of this we do not
include it in our calculation.  Using a T$_{\rm dust}$ = 36$\pm$1\ K
and using the LABOCA 870 $\micron$ flux we compute M$_{\rm dust}$ =
(4.3$\pm$0.4) $\times10^{7}\ \rm M_{\sun}$.

In order to compute the gas mass from the dust mass it is important to know
the dust-to-gas ratio (DGR). \citet{Leroy2011, Sandstrom2013} shows that there is a
correlation between metallicity and the DGR for local star forming galaxies
and parameterized it using Eq. \ref{eq:dust-to-gas} for metallicities computed
using KK04:

\begin{equation}
{\rm log}({\rm DGR}) = a + b(12 + {\rm log}({\rm O/H}) - c)
\label{eq:dust-to-gas}
\end{equation}

\noindent where a = -1.86, b = 0.85 and c = 8.39.  Using the metallicity
found for the entire galaxy from the R23 line ratio we find a DGR = 0.0114.

The relationship between molecular gas, \hi\ and dust is shown in Eq.
\ref{eq:gas_mass} \citep{Leroy2011, Sandstrom2013}

\begin{equation}
\Sigma_{\rm D}/{\rm DGR} = \Sigma_{\rm HI} + \Sigma_{\rm H_{2}}
\label{eq:gas_mass}
\end{equation}

\noindent where $\Sigma_{\rm D}$ is the mas surface density of dust,
$\Sigma_{\rm HI}$ is the mass surface density of \hi, and $\Sigma_{\rm
H_{2}}$ is the mass surface density of molecular gas. \hi\ is
difficult to detect at higher redshift \citep[e.g. $z=0.4$ galaxy 180 hours
VLA,][]{Fernandez2016} and because of this it is typically excluded from
the total gas mass.  However, LIRGs and ULIRGs (Ultra Luminous Infrared
Galaxies, L$_{\rm IR} \geq 10^{12}\ \rm L_{\sun}$) are observed in the
local Universe to have more molecular gas than atomic
\citep[e.g. H$_{2}$/\hi\ $>$ 1,][]{MirabelSanders1989}, with the ratio of
H$_{2}$/\hi\ increasing with infrared luminosity. At most, the total
molecular gas could be off by about 0.3 dex.  More typically for
submillimeter galaxies (SMGs) and ULIRGS the total \hi\ mass is not
significant \citep[0.1 dex,][]{SandersMirabel1996, Santini2010}.  We find
that the molecular gas mass is the following; $M_{\rm gas} = (3.8\pm0.4)
\times 10^{9}\ \rm M_{\sun}$.  

With the gas mass and far-IR SFR ($54\pm2\ \rm M_{\sun}\ \rm yr^{-1}$)
we compute the gas depletion timescale of 70 Myr.  If instead we assume a
more conservative SFR, from the instantaneous SFR derived from H$\alpha$
(33$\pm$1 M$_{\sun}\ \rm yr^{-1}$), it would suggest a depletion
timescale of 120 Myr. 

Comparing the gas mass to the stellar mass gives the gas fraction $f_{\rm gas}$
= 0.13$^{+0.06}_{-0.04}$, defined as $f_{\rm gas} =  M_{\rm gas}/(M_{\rm gas} +
M_{\ast})$, or $M_{\rm gas}$ = (4.1$\pm$0.4)$\times10^{9}\ M_{\sun}$.
Comparing to other galaxies at this redshift \citep{Daddi2010, Geach2011,
Magdis2012}, the gas fraction for this galaxy is typical and not unusual.
The range of gas fractions at z$\sim$0.6 is about 0.05 -- 0.3.
 
For the second method of determining the gas mass, we use the dynamical
mass of the galaxy from \S\ref{sec:kinematics}.  The gas mass is defined as
$M_{\rm gas} = M_{\rm dyn} - M_{\ast} - M_{\rm dark}$, where the $M_{\rm
dark}$ is the dark matter mass.  If we assume a dark matter fraction of 0.2
-- 0.3 within a radius of 10 kpc \citep{ForsterSchreiber2009} for galaxies
$z = 1-3$ we get a gas mass of $M_{\rm gas}$ =  (7.7 -- 12.4)$\times10^{9}\
\rm M_{\sun}$, which is $f_{\rm gas}$ = 0.23 -- 0.33, which is larger than
what we compute using the dust mass.  There are likely more uncertainties
with the dynamical mass calculation, as the dynamical mass could be
influenced by the position of the slit, not fully sampling the galaxy's
velocity field.  In addition, inclination of the galaxy could also be large
uncertainty, as the projection of the galaxy on the sky is highly
degenerate, especially when considering diverse galaxy morphologies.
Finally, the stellar mass could be underestimated, which could result in a
larger gas fraction.  From herein, we use the gas fraction based on the
dust mass.

In \S\ref{sec:ir_location} we find that as much as half of the far-IR flux
could be associated with the H {\sc ii} region (half of the SFR).  Extending
this to the molecular gas, if we assume that the molecular gas traces the
dust, then roughly half of the molecular gas could be associated with the H
{\sc ii} region.  This also suggests a similar depletion timescale of 70
Myr.  Using the instantaneous SFR derived from H$\alpha$ from the \hst\
grism ($8\ \rm M_{\sun}\ \rm yr^{-1}$) for the H {\sc ii} region would
suggest a depletion timescale of 230 Myr.  Adding the depletion
timescale to the stellar age (0.38$^{+0.31}_{-0.17}$ Gyr) of the AS1063a \hii\
region makes its lifetime 0.44 -- 0.92 Gyr.  However, even with a
Balmer decrement, it is unclear if we are truly measuring the current
instantaneous SFR or the dust attenuated one.  In order to determine the
lifetime of the \hii\ region, even without instantaneous star formation
tracers which are less affected by dust (i.e. P$\alpha$, P$\beta$), we rely
on the far-IR SFR to set the maximum rate that stars form.  We find that
incorporating the range of SFRs and stellar ages provides a lifetime for the
\hii\ region being 0.28 - 0.92 Gyr. This is consistent with \hii\ regions
at $z\sim2$ which are expected to live between 0.1 -- 1 Gyr
\citep{Dekel2009, Genzel2011, Guo2012}.

\subsection{Giant ($\sim$kpc) \hii\ Region Embedded in a Rotating Disk at $z=0.61$}

\floattable
\begin{deluxetable}{lcc} 
\tabletypesize{\footnotesize}
\tablewidth{0pc}
\tablecaption{\label{tab:phys_prop}Physical Properties of AS1063a}
\tablecolumns{3}
\tablehead{
\colhead{Parameter} & \colhead{Value} & \colhead{Unit} 
}
\startdata
T              &  36$\pm$1                         &  K \\
L$_{\rm IR}$   &  (3.1$\pm$0.1)$\times10^{11}$     &  L$_{\sun}$ \\
SFR FIR        &  54$\pm$2                         &  M$_{\sun}$/yr \\
SFR H$\alpha$  &  33$\pm$1                         &  M$_{\sun}$/yr \\
M$_{\rm dust}$ &  (4.6$\pm$0.4)$\times10^{7}$      &  M$_{\sun}$ \\
M$_{\ast}$ &  2.57$^{+1.50}_{-0.95}\times10^{10}$  &  M$_{\sun}$ \\
M$_{\rm dyn}$  &  (5$\pm$1)$\times10^{10}$         &  M$_{\sun}$ \\
M$_{\rm gas}$  &  (4.1$\pm$0.4)$\times10^{9}$      &  M$_{\sun}$ \\
E(B-V)         &  0.36$\pm$0.04                    &  \\
A$_{\rm V}$    &  1.5$\pm$0.2                    &  mag \\
f$_{\rm gas}$ (M$_{\rm dust}$)  &  0.13$^{+0.06}_{-0.04}$               & \\
i              &  61$\pm$5                             &  deg \\
\enddata

\end{deluxetable}

Taking into account the effect of magnification, we divide by the linear
magnification, determined by the lens model of the cluster, and find the
FWHM of the clump in the source plane is 996$\pm$46 pc.  In order to
compute the error bar for the size of the H {\sc ii} region we ran a Monte
Carlo simulation, varying the noise and refitting the elliptical Gaussian
for 1000 realizations, then measuring the standard deviation of the width
measurements.  

At the spatial resolution provided by the gravitational lensing, this means
that the clump appears to be a single clump, with ACS resolving individual
structures larger than 90 pc and WFC3/IR larger than 240 pc.  While we can
not completely rule out the possibility, it appears unlikely to have $>$10
smaller clumps lined up in a chain, appearing like a single 1 kpc clump.

\subsubsection{As a Low-Redshift Analog to Giant \hii\ Regions at $z \gtrsim 1$}

\begin{figure*}
\begin{center}
\includegraphics[scale=0.75]{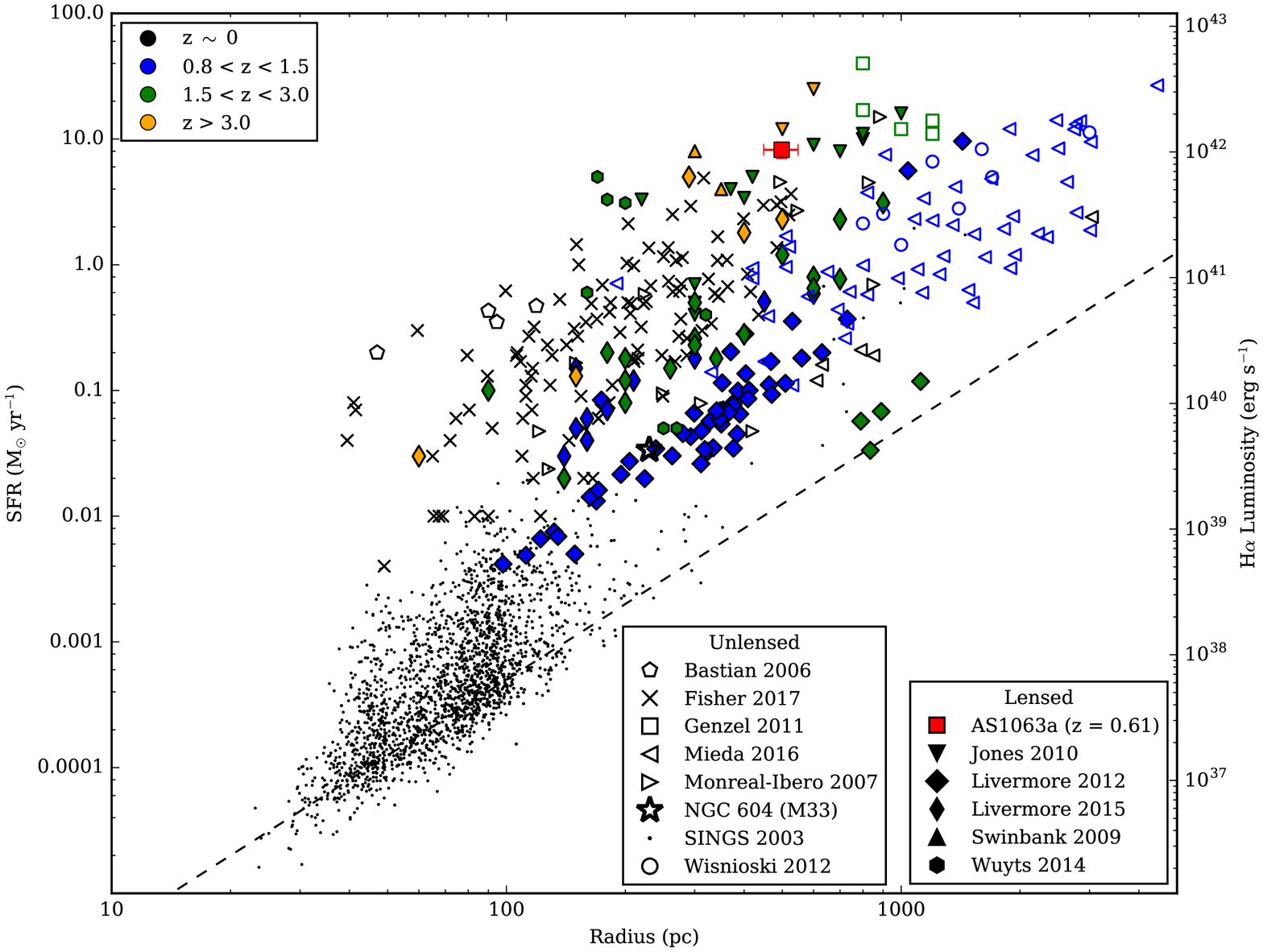}
\caption{\label{fig:HII_plot}Relation between SFR and the radius of H {\sc
ii} regions \citep[adapted from;][]{Wisnioski2012, Livermore2012}.  The red
point is the H {\sc ii} region from AS1063a at $z = 0.61$.  The other points
are all galaxies above $z>0.8$, color coded by redshift, by combining
samples of lensed \citep[solid color symbols;][]{Swinbank2009, Jones2010a,
Livermore2012, Wuyts2014, Livermore2015} and unlensed \citep[open
symbols;][]{ForsterSchreiber2011b, Wisnioski2012, Mieda2016} galaxies. The
grey points are from SINGS galaxies \citep{Kennicutt2003}.  The open black
symbols are giant H {\sc ii} regions found at $z\sim0$ in Antennae galaxies
\citep{Bastian2006} and ULIRGs \citep{Monreal-Ibero2007}.  The black star is
the giant H {\sc ii} region NGC 604 found in M33
\citep{RelanoKennicutt2009}.  The black x's are large star-forming
clumps from nearby galaxies \citep{Fisher2017}. The dashed line represents
a line of constant H$\alpha$ surface brightness, which is proportional to
R$^{2}$.}
\end{center}
\end{figure*}

Figure~\ref{fig:HII_plot} shows the SFR of individual clumps versus
their radius for nearby galaxies and galaxies with redshifts $z > 0.8$.
When plotted on this relation, the H {\sc ii} region of the $z=0.61$ lensed
galaxy found in AS1063, is more luminous than typical local clumps (by $\sim$2
orders of magnitude) and as large as the largest clumps found locally.
This H {\sc ii} region is almost a factor of 10 larger in radius than the
mean radius of local clumps.  When comparing to high redshift ($z > 0.8$)
clumps, the H {\sc ii} region is similar in size and luminosity.  What is
particularly striking about this star forming clump is how is resembles
clumps at the highest redshifts.  It appears more similar in size and
luminosity to the clumps at $z > 3.0$ (and the more luminous clumps $1.5 < z
< 3.0$). 

Local clumps, on average, are much smaller ($\sim$50-200 pc in radius)
than high redshift ($z > 0.8$) clumps and less luminous (SFR$\sim$0.0001-0.01
M$_{\sun}\ \rm yr^{-1}$).  However, there are exceptional cases of local galaxies
with large star forming regions almost spanning the size range for higher
redshift clumps, but at lower luminosity.  For comparison one of the
largest local H {\sc ii} regions is overplotted in Figure~\ref{fig:HII_plot},
NGC 604 in M33.

Local star forming galaxies (from the SINGS sample) appear to fall on a
surface brightness empirical relation, where as a clump becomes brighter,
it also gets larger.  It has been suggested that this relation evolves with
redshift in galaxies along the ``main sequence", where higher redshift
clumps are more luminous than lower redshift clumps for a given size
\citep{Livermore2012, Livermore2015}.  This evolution may be driven by a
galaxy's gas fraction where the higher redshift galaxies are found to have
higher gas fractions than the local galaxies. However, \cite{Mieda2016},
found large resolved clumps (using IFS) in high redshift ($z\sim1$)
galaxies, with wide range of SFRs, which appear to be scaled up versions of
local \hii\ regions.  Additionally, \citet{Johnson2017b} finds that for
a $z=2.48$ galaxy with a low star formation rate (9 M$_{\sun}\ \rm
yr^{-1}$) it contains star forming clumps that are 30 - 50 pc in size.   This hints that
there is a another process at work.  \citet{Cosens2018} recently suggested
that there is no redshift evolution in the current clump samples, and that there
are really two populations of star forming clumps.  The populations can be
divided based on whether or not a galaxy is undergoing a starburst (i.e.
$\Sigma_{\rm SFR}$). 

Another key feature is that the clump in AS1063a appears to have formed in
isolation (in-situ).  As we show in \S\ref{sec:kinematics} the clump
appears to be embedded in a rotating disk with no evidence of interaction.
This is typical of the clumps found at high redshift ($z>2$) but uncommon
for local galaxies. For galaxies at low redshift it is found that massive
luminous star forming regions are induced by the interacting galaxies.  It
is suggested that the conditions at higher redshift, higher gas fractions
and cold flow accretion, may be responsible for clumps forming in
isolation, perhaps due to gravitational instability. They could also be
short lived features, as work from \citet{Dekel2009}, \citet{Genzel2011}
and \citet{Guo2012} has suggested that clumps can migrate or diffuse within
0.1--1 Gyr.  We also know that H$\alpha$ emission traces recent star
formation \citep{KennicuttEvans2012} within the last 3-10 Myr.

It would be expected that such a clump at this redshift would be quite
rare. However, two of the largest and brightest clumps in the
\citet{Livermore2012} sample ($z=1-1.5$) are at redshift $z=1.01$ (A773).
These clumps are more comparable to the size and luminosity of clumps in
the \citet{Jones2010a} sample at redshift $z\sim2$ and the
\citet{Wisnioski2012} sample (WiggleZ). 
Recent work by \citet{Guo2015} has shown that a galaxies' stellar mass
determines the frequency of clumps found from redshifts $z=0.5-2$.  The
clump fraction remains constant for galaxies with a smaller stellar mass,
whereas the clump fraction decreases with time over the redshift range for
higher stellar mass galaxies.

\section{Conclusions}

In this paper we present the results of the infrared/submillimeter survey
of the core of the massive galaxy cluster AS1063.  Three bright 24
$\micron$ detected sources near (r$<$30\arcsec) the cluster core of AS1063
stand out in the survey.  Two of the sources are cluster members (AS1063b
and AS1063c) with recent star formation.  The third source is a lensed
galaxy at $z=0.61$ (AS1063a).  We also present evidence that AS1063a
contains a giant \hii\ region that is about a kiloparsec in diameter
and 2 orders of magnitude more luminous than typical local \hii\
regions, with strong Balmer line emission (H$\alpha$-H8) and forbidden line
emission (\oii\ $\lambda$3727, \oiii\ $\lambda\lambda$4959,5007, and
\neiii).

The main conclusions in this paper are the following:
\begin{itemize}

  \item We discover a gravitationally lensed galaxy (AS1063a) with a giant
luminous \hii\ region (D = 1 kpc, SFR$\sim$10 M$_{\sun}\ \rm yr^{-1}$) at
$z=0.61$.  It appears to be a single clump with no additional clumps
resolved within it larger than 90 pc (ACS resolution with the aide of
gravitational lensing).

  \item The kinematics of the \oii$\lambda3727$ doublet indicate
that the giant \hii\ region is part of a rotating disk.  There is no
evidence of any nearby galaxies interacting with AS1063a which suggests
that this \hii\ region formed in-situ.

  \item We find a significant offset between the 24 $\micron$ emission and
center of AS1063a, in which it falls between the bulge and \hii\ region of
the galaxy.  When fitting the extended 24 $\micron$ emission, we find that
it is likely that half of the star formation in AS1063a is coming from the
giant \hii\ region.  Though we cannot rule out the possibility of a
obscured star forming region, due to the degeneracy of the fits to the
MIPS PSF.

  \item Using both a dust-to-gas ratio and kinematics of the galaxy
rotation we are able to determine the gas fraction, f$_{\rm gas}$ =
0.13$^{+0.06}_{-0.04}$, which is typical for LIRGs at that redshift. At
$z\sim2$ gas fractions are normally 0.3 -- 0.8.

  \item Assuming that about half of the star formation comes from the \hii\
region and using the gas fraction, implies that the depletion timescale for
the \hii\ region is roughly between 70-230 Myr.  Incorporating the
stellar age, the lifetime of the \hii\ is predicted to be 280-920 Myr, which
makes it short-lived.

\end{itemize}

The \hii\ region in AS1063a is more luminous than local analogs and is
unexpectedly luminous for its redshift, resembling an \hii\ region at
$z\sim2$.  This could potentially be a rare occurrence, however in the
\citet{Livermore2012} sample, two star forming clumps found in a lensed
galaxy in A773, have similar luminosity and size at $z=1.01$.  In addition,
the giant \hii\ region in AS1063a appears to have formed in isolation,
and to have been not induced by a merger.  Unlike giant star forming
regions in the local Universe, the giant \hii\ region in AS1063a
appears more like the ones found in redshift $z\sim2$ galaxies.  Even
though recent studies \citep{Guo2015} have determined the fraction of
clumps in field galaxies at redshifts $z=0.5-2$, they are typically
unresolved.  Larger samples of gravitationally lensed IR galaxies will be
necessary to determine the distribution of sizes and luminosities of
resolved star forming regions.

\acknowledgments G.L.W. thanks
Dan Kelson,
Patrick Sheehan,
Justin Spilker,
Decker French,
Andrey Vayner,
Cliff Johnson, 
Maren Cosens,
Thomas Connor,
Brenda Frye,
Shelley Wright,
George Rieke,
Xiaohui Fan,
Dennis Zaritsky,
and
Dan Marrone
for useful discussions.

P.G.P.-G. acknowledges support from Spanish Government MINECO
AYA2015-70815-ERC and AYA2015-63650-P Grants. 

\facilities{APEX (LABOCA), {\it Herschel Space Observatory} (PACS,
SPIRE), {\it Hubble Space Telescope} (ACS, WFC3), Magellan:Clay
(LDSS-3, MMIRS), {\it Spitzer Space Telescope} (IRAC, IRS, MIPS)}

\software{
aXe \citep{Kummel2009},
COSMOS \citep{Dressler2011},
GALFIT \citep{Peng2002,Peng2010},
IRAF \citep{Tody1986,Tody1993},
LENSTOOL \citep{Jullo2007},
PAHFIT \citep{JDSmith2007},
SExtractor \citep{BertinArnouts1996},
TinyTim \citep{Krist2011},
UniMap \citep{Piazzo2015}}

\appendix

\section{24 $\mu$m Selected Galaxies with Spectroscopic Redshifts}
\label{sec:append}

In Table~\ref{tab:24um_src} and \ref{tab:24um_src2} we list the 71
\spitzer/MIPS 24 $\micron$ sources in this sample; 61 of which were
targeted with Magellan/LDSS-3 and the redshifts for 10 are from the
literature \cite{Gomez2012, Karman2015, Caminha2016, Rawle2016,
Karman2017,Connor2017}. Spectroscopic redshifts for all of the
\spitzer/MIPS 24 $\micron$ counterparts are listed in
Table~\ref{tab:24um_src}.  The \spitzer/MIPS 24 $\micron$ sources
targeted with LDSS-3 but a spectroscopic redshift was unable to be
determined are listed in Table~\ref{tab:24um_src2}. For the sources
targeted with LDSS-3, a spectroscopic redshift was determined with
either one or more emission or absorption features. The redshifts for
twenty-three 24 $\mu$m sources were previously unknown, mostly between
the redshifts $0.6 < z < 1.4$.  The emission lines detected are listed
in Table~\ref{tab:24um_src} for the LDSS-3 sample.  We adopt similar
spectroscopic quality codes as the DEEP2 survey.  The redshift quality
codes are defined as the following; $z_{q}$ of 2 is a single emission
or absorption feature where the redshift is dubious, $z_{q}$ of 3 there
is a few emission or absorption features where the redshift is secure,
and a $z_{q}$ of 4 several emission and absorption features and the
redshift is very secure.

\begin{figure*}
\begin{center}
\includegraphics[scale=0.7]{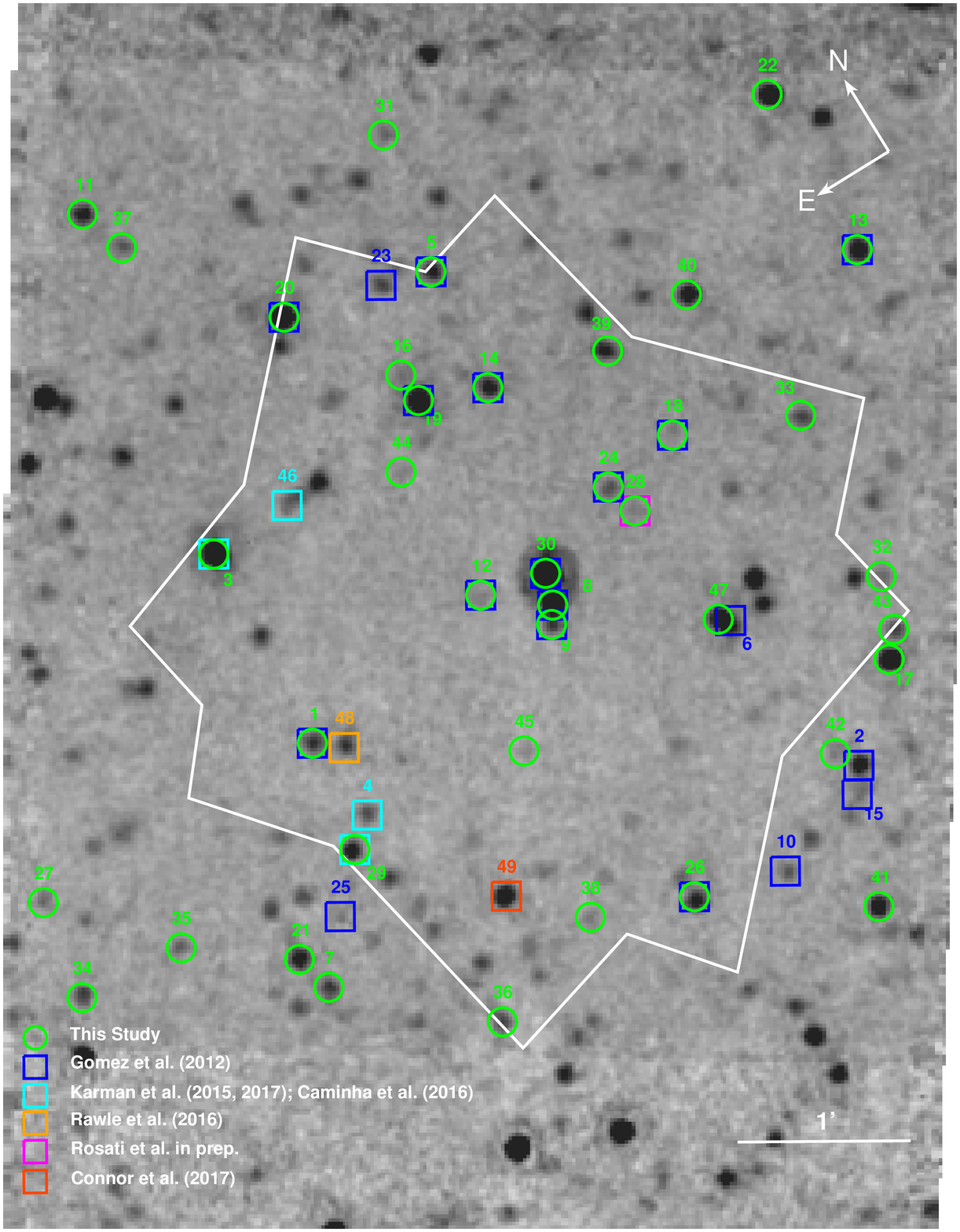}
\caption{\label{fig:24um_srcs}\spitzer/MIPS 24$\micron$ image showing
the central 5\arcmin$\times$5\arcmin\ FOV of AS1063.  All of the 24
$\micron$ sources that are marked have a redshift.  The green circles
mark the positions of 24 $\micron$ selected galaxies targeted by this
study.  The squares mark the positions of the galaxies with redshifts
from
\citet{Gomez2012,Karman2015,Caminha2016,Rawle2016,Connor2017,Karman2017}
and Rosati et al. in prep. which are now known to have 24
$\micron$ emission.  Source 30 is AS1063a (the main focus of this
paper), source 8 is AS1063b and source 9 is AS1063c. The white polygon
represents the CLASH ACS FOV.}

\end{center}
\end{figure*}

Figure~\ref{fig:24um_srcs} shows the positions of the galaxies from our
sample and the literature.  From the spectroscopic sample, we
find that 29 galaxies are in the background, 15 are cluster members and 5
are foreground galaxies.  For one of the sources the correct counterpart is
ambiguous, there are two galaxies within 1.58\arcsec\ of each other and
they both fall within the LDSS-3 slit.  Based on the distance from the
optical counterparts to the 24 $\micron$ source, the lower redshift galaxy
($z=0.474$) appears to be the more likely candidate of the 24 $\micron$
emission.  Of the 29 background galaxies; 12 have redshifts $z > 1$, one of
which is within the central arcmin of the cluster center.  Of the 24
galaxies with previously unknown redshifts, 20 of the galaxies are found in
the background of the cluster.

We also identify 5 galaxies at redshift $z\sim0.6$, one of which is
discovered in \citet{Gomez2012}.  The spatial distance between these
galaxies and AS1063a, when accounting for the deflection caused by
gravitational lensing, are 230, 470, 580 and 980 kpc.  This study,
\citet{Gomez2012} and \citet{Karman2015} identify 6 galaxies at $z\sim0.6$
all within 2.4\arcmin\ of each other and a $\Delta$v of 2400 km s$^{-1}$ (5 within
1.4\arcmin\ and with $\Delta$v of 370 km s$^{-1}$).  \citet{Gruen2013} uses weak
lensing to suggest that there is a background cluster at $z\sim0.6$, offset
by 14.5\arcmin\ from AS1063.  This could potentially be an infalling group,
however this is speculative, more wide field spectroscopy is needed to
determine the size and structure of the suspected background cluster at
$z\sim0.6$.

\floattable
\begin{deluxetable}{rccccccl}
\tabletypesize{\scriptsize}
\tablewidth{0pc}
\tablecaption{\label{tab:24um_src}Spectroscopic Redshifts of 24 $\micron$ Selected Galaxies}
\tablecolumns{8}
\tablehead{
\colhead{ID} & \colhead{R.A.} & \colhead{Decl.} & \colhead{F$_{24\micron}$(mJy)} & \colhead{z} & \colhead{z$_{q}$} & \colhead{Ref.\tablenotemark{1}} & \colhead{Emission Lines}
}
\startdata
1   & 22:48:51.250 & -44:32:04.14 & 0.222$\pm$0.010 & 0.082  & 4 & a     & [OII], H$\beta$, [OIII]$_{\lambda4959,5007}$, HeI, [OI], H$\alpha$, [NII]$_{\lambda6548,6583}$, [SII], [ArIII] \\
2   & 22:48:36.308 & -44:33:52.23 & 0.356$\pm$0.013 & 0.198  &   & a,g   & \nodata  \\
3   & 22:48:50.728 & -44:30:49.15 & 1.234$\pm$0.017 & 0.211  & 4 & b,d   & [OII], [NeIII], H$\gamma$, H$\beta$, [OIII]$_{\lambda4959,5007}$, H$\alpha$, [NII]$_{\lambda6548,6583}$, [SII] \\
4   & 22:48:50.961 & -44:32:35.86 & 0.102$\pm$0.007 & 0.237  &   & b,d,g & \nodata  \\
5   & 22:48:39.744 & -44:30:04.87 & 0.220$\pm$0.013 & 0.275  & 4 & a     & [OII], H$\beta$, [OIII]$_{\lambda4959,5007}$, H$\alpha$, [NII]$_{\lambda6548,6583}$, [SII] \\
6   & 22:48:37.389 & -44:32:45.05 & 0.725$\pm$0.016 & 0.331  &   & a,g   & \nodata \\
7   & 22:48:55.030 & -44:33:20.36 & 0.250$\pm$0.016 & 0.332  & 4 &       & [OII], H$\delta$, H$\gamma$, H$\beta$, [OIII]$_{\lambda5007}$, HeI, H$\alpha$, [NII]$_{\lambda6548,6583}$, [SII] \\
8   & 22:48:42.113 & -44:32:07.39 & 0.769$\pm$0.012 & 0.337  & 4 & a,b   & [OII], H$\beta$, [OIII]$_{\lambda4363,4959,5007}$, H$\alpha$, [NII]$_{\lambda6548,6583}$, [SII] \\
9   & 22:48:42.480 & -44:32:12.99 & 0.270$\pm$0.034 & 0.337  & 4 & a,b   & [OII], H$\beta$, [OIII]$_{\lambda5007}$, H$\alpha$, [NII]$_{\lambda6548,6583}$ \\
10  & 22:48:40.205 & -44:34:10.30 & 0.111$\pm$0.010 & 0.338  &   & a,g   & \nodata  \\
11  & 22:48:48.503 & -44:28:42.90 & 0.257$\pm$0.012 & 0.341  & 4 &       & [OII], H$\alpha$, [NII]$_{\lambda6548,6583}$, [SII] \\
12  & 22:48:43.960 & -44:31:51.02 & 0.099$\pm$0.005 & 0.347  & 4 & a,b,h & \nodata \\
13  & 22:48:27.407 & -44:31:17.28 & 0.342$\pm$0.012 & 0.348  & 4 & a     & [OII], H$\beta$, H$\alpha$, [NII]$_{\lambda6548,6583}$, [SII] \\
14  & 22:48:40.158 & -44:30:50.17 & 0.256$\pm$0.014 & 0.351  & 4 & a     & [OII], H$\beta$, [OIII]$_{\lambda5007}$, H$\alpha$, [NII]$_{\lambda6548,6583}$, [SII] \\
15  & 22:48:36.857 & -44:34:00.61 & 0.149$\pm$0.006 & 0.353  &   & a,g   & \nodata  \\
16  & 22:48:42.379 & -44:30:30.28 & 0.135$\pm$0.005 & 0.354  & 4 & h     & \nodata \\
17  & 22:48:33.617 & -44:33:26.01 & 0.511$\pm$0.012 & 0.335  & 4 &       & H$\beta$, H$\alpha$, [NII]$_{\lambda6548,6583}$, [SII] \\
18  & 22:48:35.806 & -44:31:38.61 & 0.063$\pm$0.005 & 0.335  & 4 & a     & [OIII]$_{\lambda5007}$, H$\alpha$,[NII]$_{\lambda6583}$ [SII] \\
19  & 22:48:42.325 & -44:30:41.12 & 0.801$\pm$0.017 & 0.355  & 4 & a     & [OII], [NeIII],[OIII]$_{\lambda4959,5007}$, H$\alpha$, [NII]$_{\lambda6548,6583}$, [SII] \\
20  & 22:48:44.647 & -44:29:51.22 & 0.728$\pm$0.014 & 0.355  & 4 & a     & [OII], H$\alpha$, [NII]$_{\lambda6548,6583}$, [SII] \\
21  & 22:48:55.368 & -44:33:06.54 & 0.304$\pm$0.015 & 0.403  & 4 &       & [OII], H$\beta$, [OIII]$_{\lambda4959,5007}$, H$\alpha$, [NII]$_{\lambda6548,6583}$  \\
22  & 22:48:27.245 & -44:30:13.99 & 0.406$\pm$0.011 & 0.453  & 4 &       & [OII], H$\beta$, H$\alpha$, [NII]$_{\lambda6548,6583}$, [SII] \\
23  & 22:48:41.380 & -44:29:59.64 & 0.122$\pm$0.010 & 0.454  &   & a,g   & \nodata  \\
24  & 22:48:38.502 & -44:31:42.29 & 0.128$\pm$0.012 & 0.457  & 4 & a     & [OII], H$\beta$, [OIII]$_{\lambda4959,5007}$, H$\alpha$, [NII]$_{\lambda6548,6583}$, [SII] \\
25  & 22:48:53.479 & -44:33:01.29 & 0.093$\pm$0.005 & 0.457  &   & a,g   & \nodata  \\
26  & 22:48:43.200 & -44:34:01.11 & 0.280$\pm$0.014 & 0.474  & 4 &       & [OII], H$\beta$, [OIII]$_{\lambda5007}$, H$\alpha$ \\
    &              &              & 0.280$\pm$0.014 & 0.609  & 3 & a     & H$\beta$ \\
27  & 22:49:01.556 & -44:32:01.95 & 0.049$\pm$0.005 & 0.570  & 4 &       & [OII], [OIII]$_{\lambda5007}$ \\
28  & 22:48:38.171 & -44:31:54.38 & 0.092$\pm$0.005 & 0.610  & 3 & e     & [OII] \\
29  & 22:48:51.905 & -44:32:43.89 & 0.244$\pm$0.012 & 0.610  & 4 & b,d,h & \nodata \\
30  & 22:48:41.760 & -44:31:56.53 & 2.223$\pm$0.015 & 0.611  & 4 & a,b   & [OII], [NeIII], H8, H$\epsilon$, H$\delta$, H$\gamma$, H$\beta$, [OIII]$_{\lambda4959,5007}$ \\
31  & 22:48:38.707 & -44:29:15.03 & 0.051$\pm$0.007 & 0.622  & 4 &       & [OII], [NeIII], H8, H$\epsilon$, H$\delta$, H$\gamma$, H$\beta$, [OIII]$_{\lambda4959,5007}$ \\
32  & 22:48:32.400 & -44:32:59.69 & 0.065$\pm$0.007 & 0.745  & 3 &       & [OII], [NeIII], H$\beta$ \\
33  & 22:48:31.871 & -44:31:56.48 & 0.167$\pm$0.013 & 0.746  & 4 &       & [OII], [NeIII], H$\delta$, H$\gamma$, H$\beta$, [OIII]$_{\lambda4959,5007}$ \\
34  & 22:49:02.129 & -44:32:37.61 & 0.206$\pm$0.013 & 0.746  & 3 &       & [OII], H$\beta$ \\
35  & 22:48:58.482 & -44:32:41.14 & 0.083$\pm$0.010 & 0.960  & 2 &       & [OII] \\
36  & 22:48:50.756 & -44:34:02.97 & 0.128$\pm$0.008 & 0.969  & 2 &       & [OII] \\
37  & 22:48:47.995 & -44:29:00.28 & 0.070$\pm$0.007 & 0.974  & 2 &       & [OII], H$\gamma$ \\
38  & 22:48:46.472 & -44:33:47.95 & 0.044$\pm$0.007 & 1.033  & 2 &       & [OII] \\
39  & 22:48:36.162 & -44:31:01.35 & 0.264$\pm$0.015 & 1.034  & 2 &       & [OII] \\
40  & 22:48:32.980 & -44:30:59.08 & 0.316$\pm$0.013 & 1.035  & 2 &       & [OII] \\
41  & 22:48:38.196 & -44:34:38.27 & 0.399$\pm$0.014 & 1.074  & 4 &       & [OII] \\
42  & 22:48:36.767 & -44:33:44.35 & 0.108$\pm$0.006 & 1.116  & 3 &       & [OII], [NeIII] \\
43  & 22:48:32.972 & -44:33:17.89 & 0.138$\pm$0.011 & 1.241  & 4 &       & [OII], [NeIII] \\
44  & 22:48:44.053 & -44:30:59.34 & 0.074$\pm$0.005 & 1.241  & 2 &       & [OII] \\
45  & 22:48:45.450 & -44:32:45.75 & 0.048$\pm$0.005 & 1.285  & 3 &       & [OII], H$\delta$ \\
46  & 22:48:47.830 & -44:30:48.20 & 0.191$\pm$0.005 & 1.427  &   & b,d,g & \nodata \\
47  & 22:48:37.717 & -44:32:42.35 & 0.725$\pm$0.016 & 1.440  & 4 &       & CII]$_{\lambda2327}$, [NeIV], MgII, [NeV], [OII], [NeIII] \\
48  & 22:48:50.432 & -44:32:11.41 & 0.327$\pm$0.006 & 1.440  &   & c,d,g & \nodata \\
49  & 22:48:48.543 & -44:33:25.85 & 0.348$\pm$0.011 & 2.566  &   & f,g   & \nodata \\
\enddata
\tablecomments{
Emission lines in the table are abrevations of the following:
[ArIII] - [ArIII]$\lambda7135$,
HeI - HeI$\lambda5876$,
MgII - MgII$\lambda2796,2804$,
[NeIII] - [NeIII]$\lambda3869$,
[NeIV] - [NeIV]$\lambda2425$,
[NeV] - [NeV]$\lambda3326,3346$,
[OI] - [OI]$\lambda6300$,
[OII] - [OII]$\lambda3727$,
[SII] - [SII]$\lambda6717,6731$
}
\tablenotetext{1}{(a) Spectroscopic redshift from \citet{Gomez2012}, (b) Spectroscopic redshift from \citet{Karman2015,Karman2017,Caminha2016}, (c) Spectroscopic redshift from \citet{Rawle2016}, (d) Spectroscopic
redshift from Rosati et al. in prep., (e) Spectroscopic redshift reported in Rosati et al. in prep. but not
\citet{Gomez2012} catalog, (f) Spectroscopic redshift from \citet{Connor2017}, (g) Source not targeted with LDSS-3, (h) No emission lines detected, redshift
determined with absorption features}

\end{deluxetable}

\floattable
\begin{deluxetable}{rccc}
\tabletypesize{\footnotesize}
\tablewidth{0pc}
\tablecaption{\label{tab:24um_src2}24 $\micron$ Selected Galaxies Targeted without a Redshift}
\tablecolumns{4}
\tablehead{
\colhead{ID} & \colhead{R.A.} & \colhead{Decl.} & \colhead{F$_{24\micron}$(mJy)}
}
\startdata
50  & 22:48:38.351 & -44:29:43.46 & 0.146$\pm$0.011   \\
51  & 22:48:33.829 & -44:30:30.27 & 0.083$\pm$0.009   \\
52  & 22:48:38.390 & -44:33:17.88 & 0.089$\pm$0.006   \\
53  & 22:48:51.484 & -44:32:59.22 & 0.153$\pm$0.012   \\
54  & 22:48:35.514 & -44:27:54.21 & 0.232$\pm$0.014   \\
55  & 22:48:52.765 & -44:29:30.57 & 0.650$\pm$0.011   \\
56  & 22:48:36.662 & -44:34:19.83 & 0.072$\pm$0.009   \\
57  & 22:48:38.959 & -44:34:05.20 & 0.121$\pm$0.010   \\
58  & 22:48:37.595 & -44:34:54.06 & 0.234$\pm$0.013   \\
59  & 22:48:26.323 & -44:30:46.24 & 0.077$\pm$0.005   \\
60  & 22:48:33.278 & -44:30:05.67 & 0.082$\pm$0.008   \\
61  & 22:48:31.255 & -44:30:19.07 & 0.085$\pm$0.007   \\
62  & 22:48:29.689 & -44:30:35.04 & 0.030$\pm$0.005   \\
63  & 22:48:36.515 & -44:30:14.18 & 0.073$\pm$0.009   \\
64  & 22:48:34.301 & -44:30:51.34 & 0.127$\pm$0.012   \\
65  & 22:48:40.691 & -44:30:41.67 & 0.094$\pm$0.006   \\
66  & 22:48:46.584 & -44:30:46.90 & 0.249$\pm$0.009   \\
67  & 22:48:36.050 & -44:32:36.58 & 0.432$\pm$0.012   \\
68  & 22:48:50.861 & -44:31:22.54 & 0.130$\pm$0.007   \\
69  & 22:48:53.780 & -44:32:26.19 & 0.087$\pm$0.009   \\
70  & 22:49:03.205 & -44:32:54.00 & 0.439$\pm$0.011   \\
71  & 22:48:23.267 & -44:30:38.31 & 0.078$\pm$0.006   \\
\enddata
\end{deluxetable}

\bibliography{newbib}

\end{document}